\documentclass[a4paper,11pt]{article}
\usepackage{epsfig}
\usepackage{cite}

\newcommand{\as}         {\ensuremath{\alpha_s}}

\newcommand{\asmz}       {\ensuremath{\alpha_s(M_{\mathrm{Z}})}}
\newcommand{\chisq}      {\ensuremath{\chi^2}}
\newcommand{\costt}      {\ensuremath{\cos\theta_T}}
\renewcommand{\d}          {\mathrm{d}}
\newcommand{\epem}       {\ensuremath{\mathrm{e^+e^-}}}
\newcommand{\invpb}      {\ensuremath{\mathrm{pb}^{-1}}}
\newcommand{\ksinul}     {\ensuremath{\xi_0}}
\newcommand{\ksip}       {\ensuremath{\xi_p}}
\newcommand{\mz}         {\ensuremath{M_{\mathrm{Z}}}}
\newcommand{\oa}         {\ensuremath{\mathcal{O}(\alpha_s)}}
\newcommand{\oaa}        {\ensuremath{\mathcal{O}(\alpha_s^2)}}
\newcommand{\perc}       {\%}
\newcommand{\ptin}       {\ensuremath{p_{\perp}^{\mathrm{in}}}}
\newcommand{\ptout}      {\ensuremath{p_{\perp}^{\mathrm{out}}}}
\newcommand{\qqbar}      {\ensuremath{\mathrm{q\overline{q}}}}
\newcommand{\rs}         {\ensuremath{\sqrt{s}}}
\newcommand{\rsp}        {\ensuremath{\sqrt{s'}}}
\newcommand{\rsptrue}    {\ensuremath{\sqrt{s'_{\mathrm{true}}}}}
\newcommand{\sdscd}      {\ensuremath{1/\sigma\cdot\mathrm{d}\sigma_{\mathrm{ch}}/\mathrm{d}}}
\newcommand{\stat}       {\ensuremath{\mathrm{(stat.)}}}
\newcommand{\syst}       {\ensuremath{\mathrm{(syst.)}}}
\newcommand{\tautau}     {\ensuremath{\tau^+\tau^-}}
\newcommand{\wqcd}       {\ensuremath{W_{\mathrm{QCD}}}}
\newcommand{\ww}         {\ensuremath{\mathrm{W^+W^-}}}
\newcommand{\xp}         {\ensuremath{x_p}}
\newcommand{\zg}         {\ensuremath{(\mathrm{Z}/\gamma)^{*}}}
\newcommand{\ycut}       {\ensuremath{y_{\mathrm{cut}}}}
\newcommand{\yp}         {\ensuremath{y}}
\newcommand{\zz}         {\ensuremath{\mathrm{ZZ}}}
\newcommand{\zzero}      {\ensuremath{\mathrm{Z}}}

\begin{document}

\begin{titlepage}
\begin{center}{\large   EUROPEAN ORGANIZATION FOR NUCLEAR RESEARCH
}\end{center}\bigskip
\begin{flushright}
CERN-EP/2002-057   \\ 15th July, 2002
\end{flushright}
\bigskip\bigskip\bigskip\bigskip\bigskip
\begin{center}{\huge \bf \boldmath   Charged Particle Momentum Spectra in  
\epem\ Annihilation at $\rs~=~192-209$~GeV
}\end{center}\bigskip\bigskip
\begin{center}{\LARGE The OPAL Collaboration
}\end{center}\bigskip\bigskip
\bigskip\begin{center}{\large  Abstract}\end{center}
Charged particle momentum distributions are studied in the reaction 
\epem~$\rightarrow$~hadrons, using data collected with the OPAL detector 
at centre-of-mass energies from 192~GeV to 209~GeV. The data correspond to 
an average centre-of-mass energy of 201.7~GeV and a total integrated 
luminosity of 433~\invpb. The measured distributions and derived 
quantities, in combination with corresponding results obtained at lower 
centre-of-mass energies, are compared to QCD predictions in various 
theoretical approaches to study the energy dependence of the strong 
interaction and to test QCD as the theory describing it. In general, a 
good agreement is found between the measurements and the corresponding 
QCD predictions.

\bigskip\bigskip\bigskip\bigskip
\bigskip\bigskip
\begin{center}{\large
Submitted to Eur. Phys. J.
}\end{center}
\end{titlepage}

%
%

\begin{center}{\Large        The OPAL Collaboration
}\end{center}\bigskip
\begin{center}{
G.\thinspace Abbiendi$^{  2}$,
C.\thinspace Ainsley$^{  5}$,
P.F.\thinspace {\AA}kesson$^{  3}$,
G.\thinspace Alexander$^{ 22}$,
J.\thinspace Allison$^{ 16}$,
P.\thinspace Amaral$^{  9}$, 
G.\thinspace Anagnostou$^{  1}$,
K.J.\thinspace Anderson$^{  9}$,
S.\thinspace Arcelli$^{  2}$,
S.\thinspace Asai$^{ 23}$
% ...
}\end{center}\bigskip
\bigskip
$^{  1}$School of Physics and Astronomy, University of Birmingham,
Birmingham B15 2TT, UK
\newline
$^{  2}$Dipartimento di Fisica dell' Universit\`a di Bologna and INFN,
I-40126 Bologna, Italy
\newline
$^{  3}$Physikalisches Institut, Universit\"at Bonn,
D-53115 Bonn, Germany
\newline
% ...
$^{  5}$Cavendish Laboratory, Cambridge CB3 0HE, UK
\newline
% ...
$^{  9}$Enrico Fermi Institute and Department of Physics,
University of Chicago, Chicago IL 60637, USA
\newline
% ...
$^{ 16}$Department of Physics, Schuster Laboratory, The University,
Manchester M13 9PL, UK
\newline
% ...
$^{ 22}$Department of Physics and Astronomy, Tel Aviv University,
Tel Aviv 69978, Israel
\newline
$^{ 23}$International Centre for Elementary Particle Physics and
Department of Physics, University of Tokyo, Tokyo 113-0033, and
Kobe University, Kobe 657-8501, Japan
\newline
% ...

\section{Introduction}
Charged particle momentum distributions are studied in the 
hadronic decays of virtual photons or Z bosons (hereafter 
referred to as \zg\ decays) at centre-of-mass (c.m.) energies, 
\rs, between 192 and 209 GeV, the highest available \epem\ 
energies. These distributions can be measured with relatively 
high precision. Combined with lower energy data they provide a 
powerful test of quantum chromodynamics (QCD) as the 
theory describing the strong interaction. 

Previous studies using \epem\ annihilation data at c.m. energies 
from 130 up to 189~GeV have shown that QCD based models and 
calculations give a good description of most of the measured 
distributions~\cite{OPAL-133,OPAL-161,OPAL-172-189,
ALEPH-133,L3-133-183,DELPHI-133}.
With the yet higher energy data used in this paper, we obtain
even more stringent tests of the models and theory. 
In addition we use these highest energy data in combination with 
earlier measurements at lower c.m. energies to study the energy 
dependence of the strong interaction.

We measure the charged particle momentum distribution, $\sdscd\xp$, 
and the \ksip\ distribution, $\sdscd\ksip$, where $\xp=2p/\rs$, 
$\ksip=\ln(1/\xp)$, $p$ is the particle momentum, $\sigma$ is 
the cross-section for non-radiative hadronic \zg\ decays and 
$\sigma_{\mathrm{ch}}$ is the charged particle cross-section in 
these events. We also measure the distribution of the rapidity, 
$y= \frac{1}{2}\left|\ln\left({{E+p_\Vert}\over{E-p_\Vert}}\right)\right|$, 
where $p_\Vert$ is the momentum component parallel to the 
thrust axis and $E$ is the energy of the particle, and the 
distributions of the 3-momentum components in, \ptin, and 
perpendicular to, \ptout, the event plane. This plane is defined 
by the eigenvectors associated with the two largest eigenvalues 
of the momentum tensor, as in reference~\cite{OPAL-133}. 

We present comparisons of the fractional momentum distributions 
$\sdscd\xp$ and $\sdscd\ksip$ in the range $\rs=14-202$~GeV to QCD 
predictions in different theoretical approaches. As in earlier 
publications~\cite{OPAL-133,OPAL-161,OPAL-172-189} we also study the 
evolution with the c.m. energy of the peak 
position \ksinul\ in the \ksip\ distribution.

In Section~\ref{section:detector} a brief description of the OPAL 
detector is given. The samples of data and simulated events used 
in the analysis are described in Section~\ref{section:samples}. In 
Section~\ref{section:analysis} the event selection and analysis procedure 
are described. In Section~\ref{section:results} we present the observed 
distributions and, together with those obtained at lower energies, 
compare them to the predictions of different Monte Carlo programs 
and analytic QCD predictions.
A summary and conclusions are given in Section~\ref{section:summary}.

\section{ The OPAL detector}
\label{section:detector}
The OPAL detector was operated at the LEP \epem\ collider at CERN. 
A detailed description can be found in~\cite{opaldetector}. 
The analysis presented here relies mainly on the measurement of 
momenta and directions of charged particles in the tracking chambers 
and of energy deposited in the electromagnetic calorimeters of the 
detector.
 
All tracking systems were located inside a solenoidal magnet which
provided a uniform axial magnetic field of 0.435~T along the beam
axis\footnote{In the OPAL coordinate system the $z$ axis points in the 
direction of the electron beam, $\theta$ is the polar angle with respect 
to this axis and $\phi$ is the azimuthal angle.}.
The magnet was surrounded by a lead glass electromagnetic
calorimeter and a hadron calorimeter of the sampling type.  Outside
the hadron calorimeter, the detector was surrounded by a system of muon
chambers.  There were similar layers of detectors in the forward and
backward endcaps.

The tracking system consisted of a silicon microvertex detector,
an inner vertex chamber, a large volume jet chamber,
and specialised chambers at the outer radius of the jet chamber to 
improve the measurements in the $z$ direction.
The main tracking detector was the central jet chamber. This device was
approximately 4~m long and had an outer radius of about 1.85~m. It had 24
sectors with radial planes of 159 sense wires spaced by 1~cm. The
momenta $p$ of tracks in the $r$-$\phi$ plane were measured with a precision
$\sigma_p/p=\sqrt{0.02^2+(0.0015\cdot p[\mathrm{GeV}/c])^2}$.
 
The electromagnetic calorimeters in the barrel and the endcap sections
of the detector consisted of 11704 lead glass blocks with a depth of
$24.6$ radiation lengths in the barrel and typically $22$ radiation lengths 
in the endcaps. The barrel section covered the angular region 
$\vert\cos{\theta}\vert<0.82$ and the endcap section the region 
$0.82<\vert\cos{\theta}\vert<0.98$.

\section{ Data and Monte Carlo samples }
\label{section:samples}
\begin{figure}[t]
\begin{center}
\resizebox{15cm}{!}
{\includegraphics{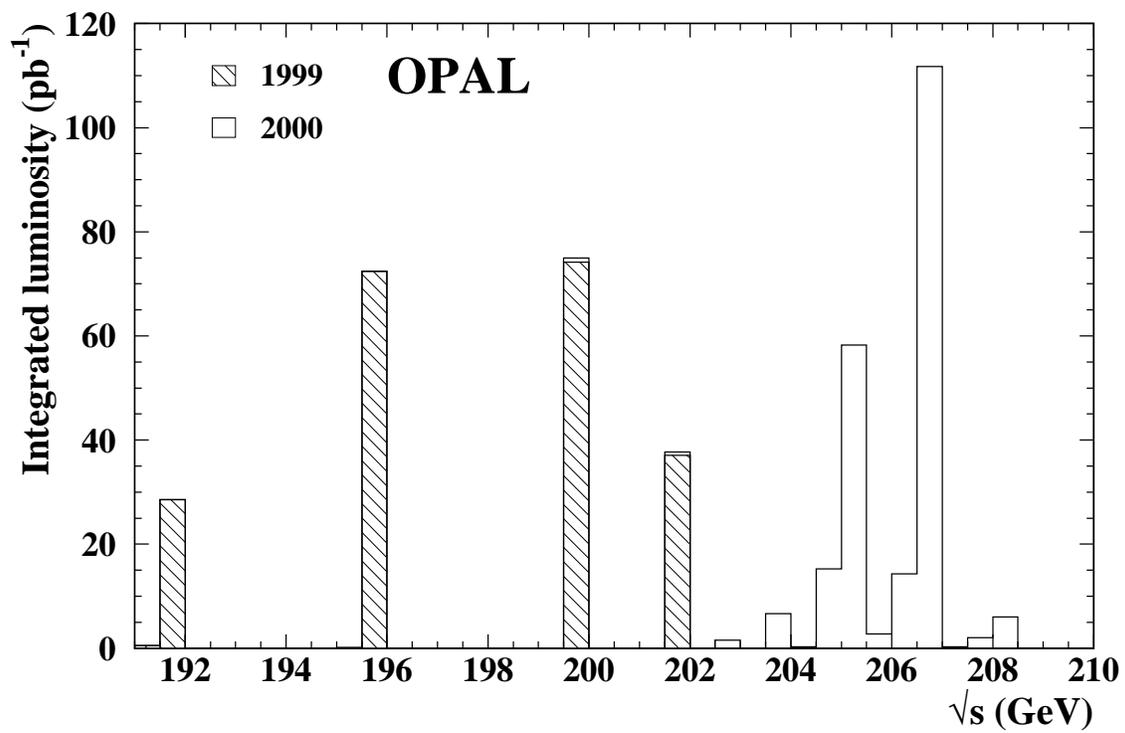}}
\caption[]{ Integrated luminosity collected by OPAL in 1999 and 2000 
at different c.m. energies.}
\label{fig_lumi}
\end{center}
\end{figure}

The data used in this measurement were recorded in 1999 and 2000, the 
final two years of the LEP~2 program. During this time \epem\ 
interactions were collected at c.m. energies in the range 192-209~GeV.
The total integrated luminosity of the data sample, as evaluated using 
small angle Bhabha collisions, is 433~\invpb. 
The luminosity collected at the different c.m. energies is 
detailed in Figure~\ref{fig_lumi} and in Table~\ref{tab:select}.

Monte Carlo event samples were generated at c.m. energies of 
192, 196, 200, 202, 206 and 208~GeV and were processed using a full 
simulation of the OPAL detector~\cite{gopal}. Distributions at 
intermediate energies were obtained by linear interpolation between 
the distributions at the nearest neighbouring available c.m. energies.
In general, the Monte Carlo samples 
contain at least 10 times the statistics of the data. 
Signal $\epem\rightarrow\zg\rightarrow\qqbar({\rm g})$ events were 
generated using the PYTHIA~6.125~\cite{jetset74,pythia61} Monte Carlo 
program for the parton shower and hadronisation stage, which was 
interfaced with the KK2F program~\cite{kk2f} to obtain a more accurate 
description of initial state radiation (ISR). The simulation parameters 
were tuned to OPAL data taken at the \zzero\ peak~\cite{opaltune}.
The PYTHIA Monte Carlo model has previously been shown to provide a 
good description of \epem\ annihilation data for c.m. energies up 
to 189~GeV (see e.g. \cite{OPAL-133,OPAL-161,OPAL-172-189,ALEPH-133,
L3-133-183,DELPHI-133,opaltune}).  

As an alternative to the string fragmentation model implemented in 
PYTHIA, events were generated using the HERWIG~5.9~\cite{herwig} 
Monte Carlo program, also tuned to OPAL data as described 
in~\cite{OPAL-161,opaltune}. The HERWIG Monte Carlo 
program implements the cluster fragmentation model.

In addition, we generated events of the type $\epem\rightarrow 4$
fermions (diagrams without intermediate gluons).  These 4-fermion
events, in particular those with four quarks in the final state,
constitute the major background in this analysis.  
Simulated 4-fermion events with hadronic and leptonic final
states were generated using the GRC4F~2.1~\cite{grc4f} Monte Carlo
program.  The final states were produced via $s$-channel or
$t$-channel diagrams, including \ww\ and \zz\ production.
This generator was interfaced to JETSET~7.4~\cite{jetset74} 
using the same parameter set~\cite{opaltune} for the 
fragmentation and decays as used for \zg\ events. 
The JETSET program implements the string fragmentation model.

Two other possible sources of background events were simulated. 
Hadronic two-photon processes were evaluated using the 
VERMASEREN~\cite{vermaseren}, HERWIG and PHOJET~\cite{PHOJET} 
programs. 
Production of $\epem\rightarrow\zg\rightarrow\tautau$ was evaluated 
using the KK2F event generator in combination with the 
TAUOLA~\cite{Tauola} decay library to simulate the decay of the 
tau leptons.

In addition to PYTHIA and HERWIG we also used the 
ARIADNE~4.08~\cite{ariadne} event generator to compare to our 
corrected data distributions. The ARIADNE program implements the 
colour dipole model to describe the parton shower process. For 
the fragmentation stage the generator was interfaced to the 
JETSET~7.4 program. The parameter set used for ARIADNE is 
documented in \cite{artune}. This model provides a good description 
of global \epem\ event properties at $\rs=\mz$, as do PYTHIA 
and HERWIG.

\section{ Data analysis }
\label{section:analysis}
\subsection{ Selection of events }
The majority of hadronic events produced at c.m. energies above 
the \zzero\ resonance are radiative events in which initial state 
photon radiation reduces the invariant mass of the hadronic system 
to about \mz. An experimental separation between radiative and 
non-radiative events is therefore required. In addition, at the 
present energies above 190~GeV, \zz\ and \ww\ production are 
kinematically possible and form a significant background to the 
$\zg\rightarrow\qqbar$ events. In this paper we use similar 
techniques to those of our previous analyses of \epem\ 
annihilation data at lower c.m. 
energies~\cite{OPAL-133,OPAL-161,OPAL-172-189}, to select 
non-radiative $\zg\rightarrow\qqbar$ events.

\subsubsection{ Preselection }
Hadronic events are identified using criteria as described
in~\cite{TKMH}. The efficiency of selecting non-radiative 
hadronic events is greater than 98\perc\ at all c.m. energies, 
as can be seen in Table~\ref{tab:select}. We define as particles 
tracks recorded in the tracking chambers and clusters recorded 
in the electromagnetic calorimeter. The tracks are required to 
have transverse momentum relative to the beam axis, 
$p_T > 150$~MeV/$c$, a number of hits in the jet chamber, 
$N_{\rm hits} \geq 40$, a distance of the point of closest approach to 
the collision point in the $r$-$\phi$ plane, $d_0 \leq 2$ cm, and 
along the $z$ axis, $z_0 \leq 25$ cm. The clusters in the 
electromagnetic calorimeter are required to have a minimum energy 
of 100~MeV in the barrel and 250~MeV in the endcap sections.

\subsubsection{ ISR-fit selection }
To reject radiative events, we determine the effective c.m. energy 
\rsp\ of the observed hadronic system as follows~\cite{OPALPR183}. 

First isolated photons are identified by looking for energy deposits 
greater than 3~GeV in the electromagnetic calorimeter, having less than
1~GeV of additional energy deposited in a cone of 0.2 radians around its 
direction. The remaining particles are formed into jets using the 
Durham~\cite{durham} algorithm with a value for the resolution 
parameter $\ycut=0.02$. A matching algorithm~\cite{MT} is employed 
to reduce double counting of energy in cases where charged tracks 
point towards electromagnetic clusters. The energy of additional 
photons emitted close to the beam direction is estimated by 
performing three separate kinematic fits assuming zero, 
one or two such photons. Of the acceptable fits, the one with the 
lowest number of photons is selected. The value of 
$\rsp$ is computed from the fitted momenta of the jets, excluding 
photons identified in the detector or close to the beam directions.
The value of \rsp\ is set to \rs\ if the fit assuming zero initial
state photons was selected. The 4-momenta of all measured particles
are boosted into the rest frame of the observed hadronic system.
 
\begin{figure}[t]
\begin{center}
\resizebox{15cm}{!}
{\includegraphics{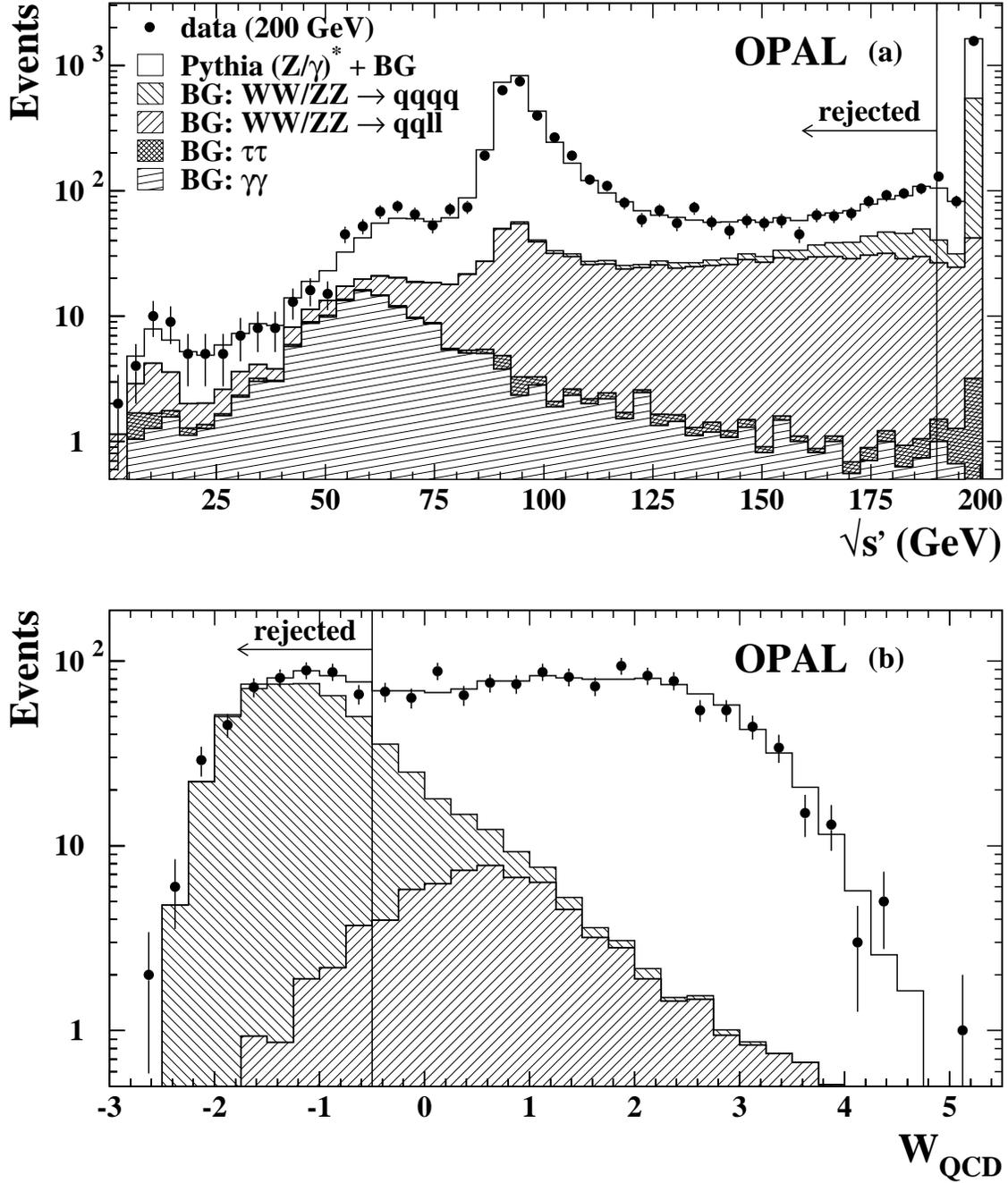}}
\caption[]{ Distributions of (a) the effective c.m. energy \rsp\ 
and (b) the QCD event weight \wqcd\ in the data at $\rs=200$~GeV 
(full points). The data are shown with statistical errors only. 
Estimated signal and background (BG) contributions are shown as 
detailed in the upper figure. The vertical lines indicate where 
the selection cuts are applied.  }
\label{fig_1}
\end{center}
\end{figure}

To reject events with large initial-state radiation (ISR), we require
$\rsp>\rs-10$~GeV. This is referred to as the ``ISR-fit'' selection.
In Figure~\ref{fig_1}a we compare the $\rsp$ distribution of the data 
set at \rs=200~GeV, after the preselection was applied, to simulated 
\zg\ events and to 4-fermion and other 
background events.  Simulated \zg\ events are classified into 
radiative events, $\rsptrue < \rs - 1$~GeV, where \rsptrue\ is the true 
effective c.m. energy, and non-radiative events, which is the complement. 
Though about 27\perc\ of the selected \zg\ events are by this 
definition radiative events, the fraction of selected events with 
$\rsptrue < \rs - 10$~GeV is only 5\perc. The 
background from 4-fermion events and the efficiency of selecting 
non-radiative events are given in Table~\ref{tab:select}.  

\subsubsection{ Final selection }
The estimated background from $\epem\rightarrow\tautau$ and two-photon 
events of the type $\gamma\gamma\rightarrow\qqbar$ is small at high \rsp.
To reject this background further and to ensure events are well 
contained in the OPAL detector we require at least seven accepted 
tracks, and the cosine of the polar angle of the thrust axis, 
$|\costt|<0.9$. The background from the $\tautau$ and two-photon 
events in the final selected sample is estimated from 
Monte Carlo samples to be about 0.1\perc\ and is neglected.

To reduce the background of 4-fermion events in the remaining
sample, we test the compatibility of the events with QCD-like
production processes. A QCD event weight \wqcd\ is computed 
as follows.  We force each event into a four-jet configuration in the
Durham jet scheme and use the EVENT2~\cite{event2} program to
calculate the \oaa\ matrix element $\left| {\cal M}
(p_1,p_2,p_3,p_4) \right|^2 $ for the processes $\epem\rightarrow
\mathrm{q\bar{q}q\bar{q},q\bar{q}gg}$~\cite{ERT}, where 
$p_1$, $p_2$, $p_3$ and $p_4$ are the momenta of the reconstructed jets. 
Since neither quark nor gluon identification 
is performed on the jets, we calculate the matrix element for each 
permutation of the jet momenta and use the permutation with the largest 
value for the matrix element to define the event weight,
\begin{equation}
\wqcd = \max_{\left\{ p_1,p_2,p_3,p_4 \right\}} \log\left( \left| {\cal M}
(p_1,p_2,p_3,p_4) \right|^2 \right).
\end{equation}
Note that the definition of the event weight contains kinematic 
information only and is independent of the value of \as.

The weight \wqcd\ is expected to have large values for
processes described by the QCD matrix element, originating from
$\zg\rightarrow\qqbar$, and smaller values for \ww\ or \zz\ events. 
In Figure~\ref{fig_1}b we compare the data distribution of \wqcd, after 
the ISR-fit selection, to the expectations of our simulation. 
A good separation between the \zg\ and 4-fermion events is achieved 
by requiring $\wqcd \geq -0.5$.

\begin{figure}[t]
\begin{center}
\resizebox{15cm}{!}
{\includegraphics{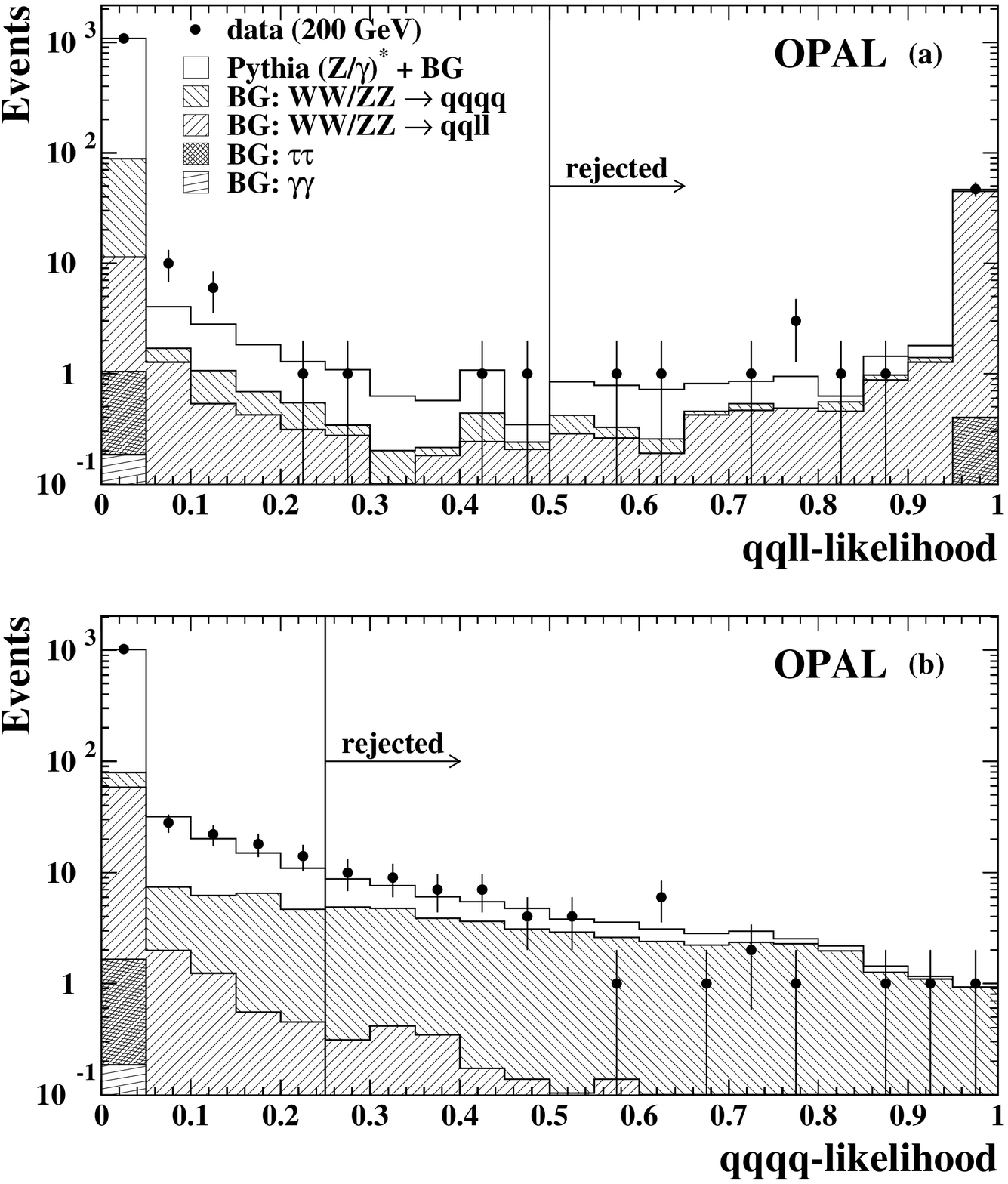}}
\caption[]{ Distributions of the likelihood for events to be (a) 
a semi-leptonic 4-fermion event or (b) a hadronic 4-fermion event, 
in the data at $\rs=200$~GeV (full points). The data are shown 
with statistical errors only. Estimated signal and background 
(BG) contributions are shown as detailed in the upper figure. 
The vertical lines indicate where the selection cuts are applied.  }
\label{fig_2}
\end{center}
\end{figure}

In addition to the cut on \wqcd, used also in our earlier papers, we  
apply a likelihood based rejection of hadronic and semi-leptonic 
\zz\ and \ww\ decays (see~\cite{OPALPR321} and references therein). 
The distributions for the semi-leptonic and hadronic likelihoods, 
after all selection cuts described above, are shown in 
Figure~\ref{fig_2}a and~\ref{fig_2}b, for the data and for the 
signal and background Monte Carlo expectations. Events are rejected 
if the semi-leptonic likelihood is greater than 0.5 or the hadronic 
likelihood is greater than 0.25, thus reducing the remaining 
4-fermion background by more than a factor two while reducing the 
expected signal by only about 2.2\perc. After the final selection 
criteria the estimated efficiency for selecting signal events is 
around 76\% while the remaining 4-fermion background in estimated 
to be around 5\%. The individual numbers for each of the six energy 
ranges are given Table~\ref{tab:select}.

\subsection{ Correction procedure }
The remaining 4-fermion background in each bin of each observable 
was estimated by 
Monte Carlo simulation and is subtracted from the observed bin content.  
A bin-by-bin multiplication procedure is then used to correct the
observed distributions for the effects of detector resolution and
acceptance as well as for the presence of remaining radiative \zg\
events.  
A bin-by-bin correction procedure is suitable for the measured 
distributions as the effects of finite resolution and acceptance  
cause only limited migration (and therefore correlation) between bins and 
the data are well described by the Monte Carlo programs used to determine 
the corrections.

For the multiplicative correction each bin, after 
background subtraction, is corrected from the ``detector level'' to 
the ``hadron level'' using two samples of Monte Carlo \zg\ events at 
each c.m. energy.  The hadron level sample does not include initial 
state radiation or detector
effects and allows all particles with lifetimes shorter than $3\times
10^{-10}$~s to decay.  The detector level sample includes full simulation 
of the OPAL detector and initial state radiation and contains only those
events which pass the same cuts as are applied to the data.  The
bin-by-bin correction factors are derived from the ratio of the
distributions at the hadron level to those at the detector level. The 
correction factors tend to be around 1.15 in most bins, with values
further away from 1 at the edges of the distributions.

\subsection{ Systematic uncertainties }
\label{section:systematics}
The experimental systematic uncertainties are estimated by 
repeating the analysis with varied experimental conditions.
To reduce statistical fluctuations in the magnitudes of the 
systematic uncertainties the average over three neighbouring 
bins is taken. In addition, all results and errors obtained 
for the different energy ranges are combined in a single set 
of results, as described in Section~\ref{section:combination}, 
thus further reducing any statistical fluctuations in the 
systematic uncertainties.

Possible inadequacies in the simulation of the response of the detector 
in the endcap regions are accounted for by restricting the analysis to 
the barrel region of the detector, requiring the thrust axis of accepted 
events to lie within the range $|\costt|<0.7$. This reduces the event 
sample by approximately 26\perc.  
The corresponding systematic error is the deviation of the results from 
those of the standard analysis.

To evaluate uncertainties due to inadequacies in the track modelling
the selection criteria for tracks were modified.
The maximum allowed distance of the point of closest approach of a track 
to the collision point in the $r$-$\phi$ plane, $d_0$, was changed from 
2 to 5~cm, the maximum distance in the $z$ direction, $z_0$, from 25 
to 10 cm and the minimum number of hits from 40 to 80. These 
modifications change the number of selected tracks by up to approximately 
12\perc. In addition, track momenta in the Monte Carlo samples were 
smeared to degrade the resolution by 10\perc. The quadratic sum over 
the deviations from the standard result, obtained from each of these 
variations, is included in the total systematic uncertainty.

Uncertainties arising from the selection of non-radiative events are
estimated by repeating the analysis using a different
technique\cite{OPAL-161} to determine the value for \rsp.  This
technique differs from our standard \rsp\ algorithm in that in this case
the kinematic fit always assumes one photon, either unobserved close
to the beam direction, or detected in the ECAL.
The final event sample with this \rsp\
algorithm has an overlap of approximately 97\perc\ with the standard
sample and is approximately the same size.  
The difference relative to the standard result is included in the total 
systematic error.

Systematic uncertainties associated with the subtraction of the
4-fermion background events are estimated by varying the cut on \wqcd\ 
and on the 4-fermion likelihoods.
We use $\wqcd\geq-0.8$ which increases the event sample 
by approximately 2\perc, and $\wqcd\geq 0$ which reduces the event sample 
by approximately 8\perc. We also vary the cuts on the likelihood values 
between 0.1 and 0.4 for hadronic 4-fermion events and between 0.25 and 0.75 
for semi-leptonic events.
The maximum deviation from the standard result obtained from the variations 
of the cuts on \wqcd\ and the hadronic 4-fermion likelihood is included in 
the total systematic uncertainty, as is the maximum deviation obtained from 
the variations of the cut on the semi-leptonic likelihood. 

In addition, we vary the predicted background to be subtracted,
up and down by 5\perc, slightly more than its measured uncertainty 
at \rs=~189~GeV of 4\perc~\cite{OPALPR321}, and include the largest 
difference from the standard result in the systematic uncertainty.

The difference in the results when we determine the bin-to-bin correction 
factors from simulated \zg\ events
generated using HERWIG instead of PYTHIA is taken as the uncertainty
in the modelling of the \zg\ events.

All systematic uncertainties are added in quadrature to obtain the total 
systematic errors. The largest contribution to the total uncertainty is that 
obtained using HERWIG as an alternative hadronisation model in the 
correction procedure.
Other significant contributions to the systematic uncertainties arise from 
varying the quality criteria on tracks and from varying the 4-fermion 
rejection cuts. 

\subsection{ Combining results }
\label{section:combination}
The event selection, the correction procedure and the estimate of 
systematic uncertainties are performed independently for each of the six 
energy regions detailed in Table~\ref{tab:select}. Afterwards the results 
are combined to obtain a single set of results with the best possible 
statistical precision. The combination procedure assumes all systematic 
uncertainties to be fully correlated between the energy regions. This means 
data points obtained at different energies are combined into a weighted 
average, using the statistical errors to determine the weights. 

To obtain the systematic uncertainty on the combined results all 
individual contributions to the systematic uncertainty, described in 
Section~\ref{section:systematics}, are first determined separately 
in each energy region. For the combined results the individual 
contributions to the systematic uncertainty are then taken as the 
average over the corresponding values obtained in each of the six 
energy regions. Finally, the combination of the thus averaged 
individual contributions into the total systematic uncertainty 
on the combined results is done, as in 
Section~\ref{section:systematics}, by adding all contributions in 
quadrature.

\section{ Results and comparison to theory}
\label{section:results}
\begin{figure}[t]
\begin{center}
\resizebox{15cm}{!}
{\includegraphics{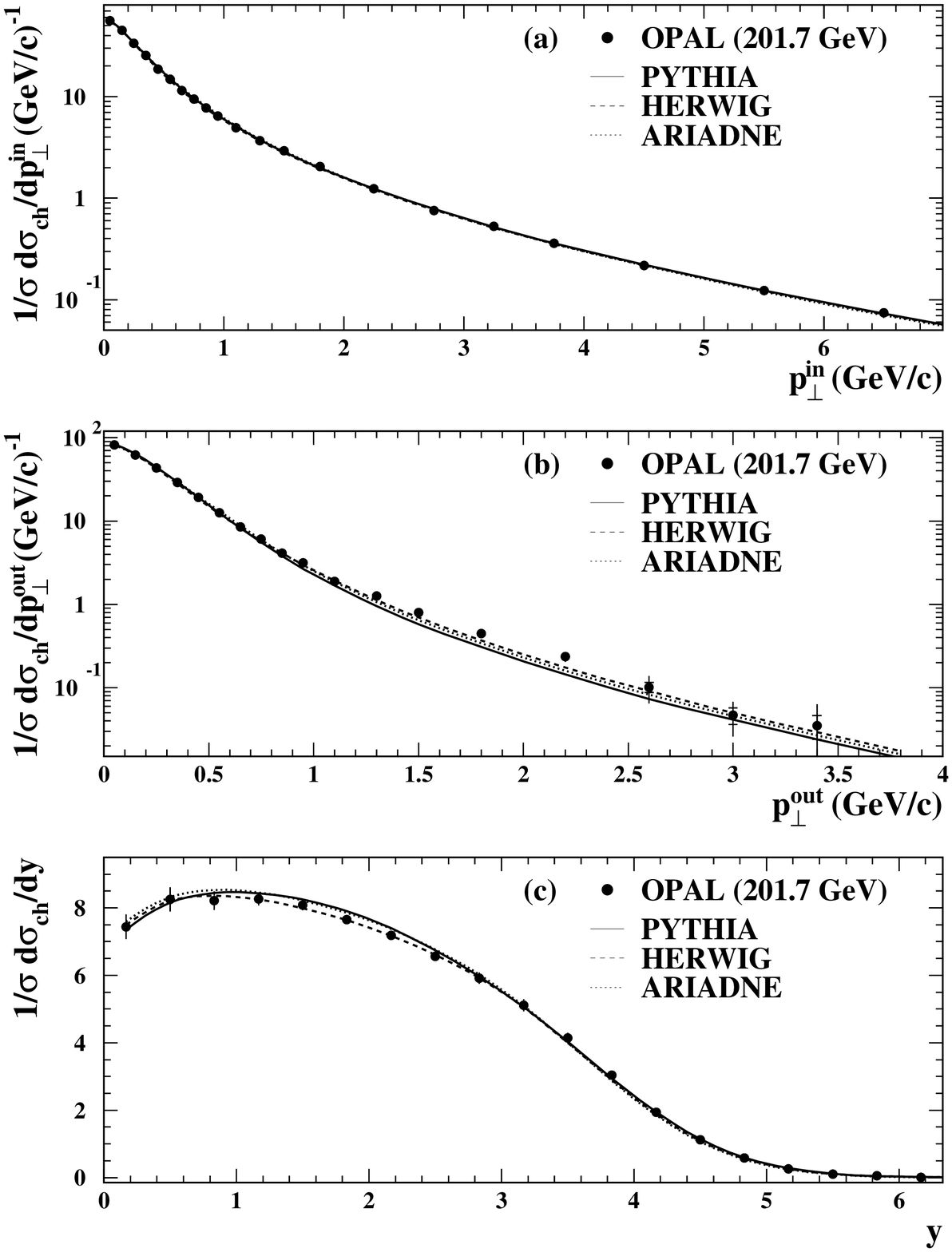}}
\caption[]{ Distributions of the charged particle momenta (a) 
\ptin\ and (b) \ptout\ and of the rapidity (c) \yp\ at 
$\langle\rs\,\rangle=201.7$~GeV, compared to the corresponding 
PYTHIA, HERWIG and ARIADNE predictions. The statistical errors 
and the combined statistical and systematic errors are shown as 
the inner and outer error bars, respectively.  }
\label{fig_dist1a}
\end{center}
\end{figure}

\begin{figure}[t]
\begin{center}
\resizebox{15cm}{!}
{\includegraphics{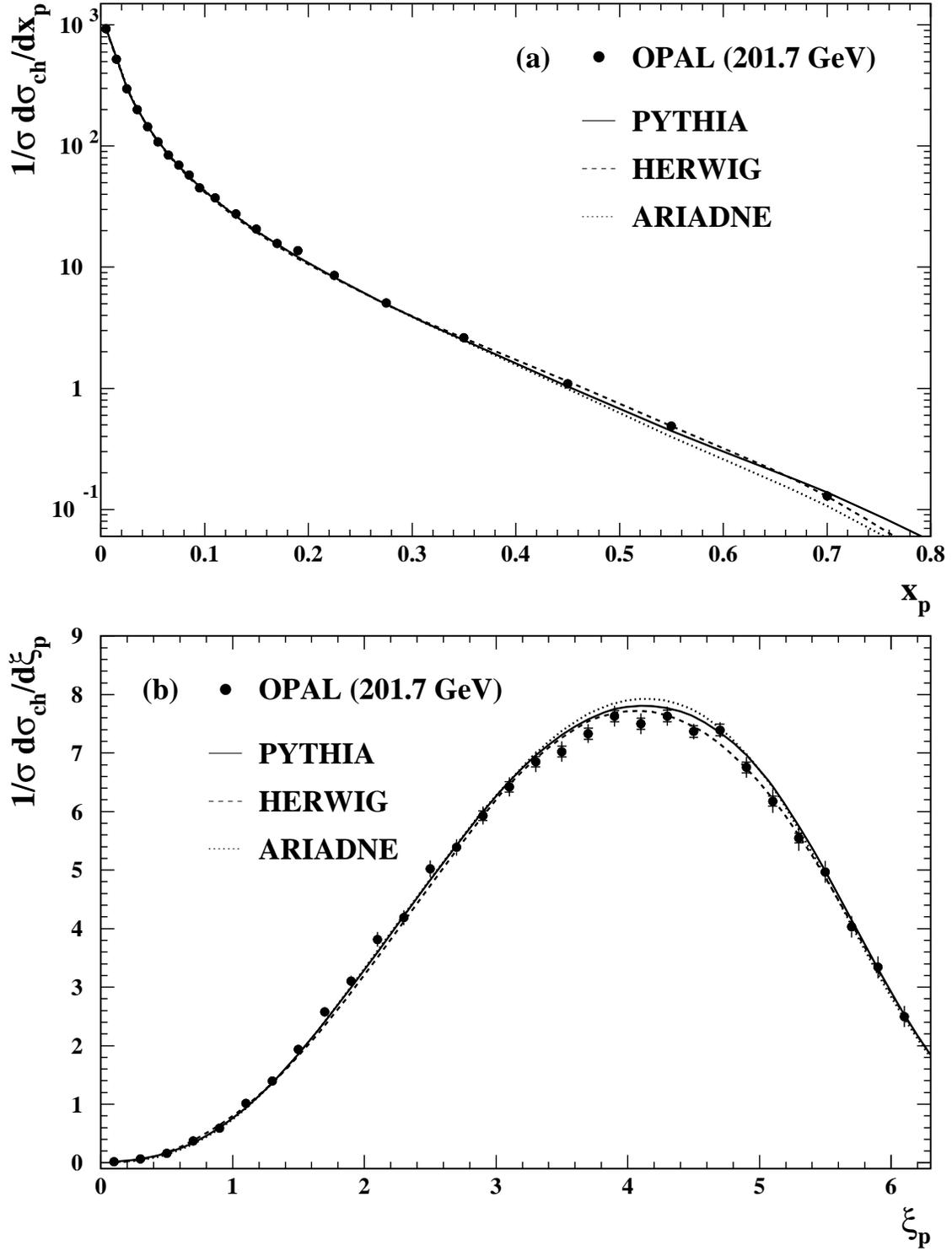}}
\caption[]{ Distributions of the charged particle fractional 
momenta (a) \xp\ and (b) \ksip\ at $\langle\rs\,\rangle=201.7$~GeV, 
compared to the corresponding PYTHIA, HERWIG and ARIADNE predictions. 
The statistical errors and the combined statistical and systematic 
errors are shown as the inner and outer error bars, respectively. }
\label{fig_dist1b}
\end{center}
\end{figure}

The \ptin, \ptout, \yp, \xp\ and \ksip\ distributions, averaged over 
the full c.m. energy range $\rs~=~192-209$~GeV, are tabulated in 
Tables~\ref{tab:pi} to \ref{tab:xi} and shown in 
Figures~\ref{fig_dist1a} and \ref{fig_dist1b}. 
These distributions correspond to an average c.m. energy of 
$\langle\rs\,\rangle=201.7$ GeV with a standard deviation of 4.8~GeV. 
They are compared to the corresponding predictions 
of the PYTHIA, HERWIG and ARIADNE Monte Carlo programs. In general, a 
reasonable agreement is observed between the predictions and the data for 
the different distributions. The HERWIG Monte Carlo program agrees  
better with the data than the PYTHIA and ARIADNE models. 
The \ptout\ distribution is harder in the data than predicted by 
any of the Monte Carlo programs. 

There are different theoretical approaches in which fractional 
energy\footnote{The $x$ and $\xi$ variables used in the description 
of the theory predictions are defined equivalently to \xp\ and \ksip, 
but refer to the fractional energies rather than the fractional 
momenta of particles.} spectra of final state hadrons can be calculated. 
At high $x$, conventional perturbation theory, utilising expansions in 
powers of \as, is applicable. Non-perturbative effects occurring at 
the final stage of the hadronisation process are not calculable but 
they can be factorised out and left to be determined experimentally. 
At low $x$, perturbation theory can also be applied, provided that 
logarithmically enhanced terms (e.g. $\sim\ln(1/x)$) are resummed 
to all orders. Remaining singular effects are not accounted for and 
as a consequence the predictions obtained are formally only valid for 
asymptotically high energies. Corrections to these asymptotic 
predictions can be calculated as an expansion in powers of $\sqrt{\as}$. 
An important precondition to the validity of these asymptotic 
predictions is that the non-perturbative hadron formation process is 
local, i.e. the properties of the hadronic final state closely follow 
the properties of the partons before hadronisation. This presumed 
equivalence is referred to as Local Parton-Hadron Duality 
(LPHD)~\cite{lphd}. For a recent review of particle production in the 
low $x$ region we refer to~\cite{khozeochs}.

In the following, we compare \xp\ and \ksip\ distributions over a large 
range of c.m. energies to theoretical predictions applicable to either 
the low $x$ or high $x$ regions, in order to test the validity of 
these QCD approaches and the LPHD hypothesis.
Since the ranges in \xp\ or \ksip\ over which the different predictions 
are valid depend not only on the terms included in 
these predictions, but also on those that are missing, we do not 
quantitatively know these ranges a priori.
Our approach, therefore, is to fit predictions to the data in a range where 
they can be observed to describe the data well. We subsequently vary these 
ranges and extrapolate the fitted predictions outside them to establish the 
limits of the validity of the predictions. 

\subsection{\boldmath Comparisons to theory at low $x$}\label{section:low-x}
\subsubsection{Fong and Webber prediction}\label{section:fw}
At low $x$, QCD predicts destructive interference effects in soft gluon 
emissions, causing the asymptotic shape of the $\xi$ distribution to be 
Gaussian~\cite{dfk-coh}. The position of the peak in the distribution is 
predicted to increase linearly with $\tau=\ln{(\rs/2\Lambda)}$, 
where $\Lambda$ is the QCD scale, defined using the one loop expression: 
 $\as(\tau)=2\pi/\beta\tau$, with $\beta=11-2N_f/N_c$, $N_c$ the number 
of colours and $N_f$ the number of active flavours. 
Corrections to this asymptotic prediction were calculated up to \oa\ by 
Fong and Webber~\cite{fongwebber91}, yielding a skewed Gaussian shape 
for the $\xi$ spectrum:
\begin{equation}
F_q(\xi,\tau) = {N(\tau)\over \sigma \sqrt{2\pi}} 
                \exp\left(   {k\over 8} 
                           - {s\delta\over 2} 
                           - {(2+k)\delta^2\over 4} 
                           + {s\delta^3\over 6} 
                           + {k\delta^4\over 24}
                    \right),
\label{eq:fw}
\end{equation}
where $\delta=(\xi-\bar{\xi})/\sigma$ and, for quark initiated jets:
\begin{eqnarray}
\bar{\xi} &=& 
{\tau\over 2} \left( 1 + {\rho\over 24}\sqrt{48\over \beta\tau}
              \right)
              \left( 1 - {\omega\over 6\tau}
              \right)  
             + \bar{\xi}_0  \,,\\
\sigma &=& 
\sqrt{\tau\over 3} \left( {\beta\tau\over 48}  
                   \right)^{1\over 4} 
                   \left( 1 - {\beta\over 64}\sqrt{48\over \beta\tau}  
                   \right)
                   \left( 1 + {\omega\over 8\tau}   
                   \right)    \,,\\
s &=& 
-{\rho\over 16} \sqrt{3\over \tau} \left( {48\over\beta\tau}  
                                   \right)^{1\over 4}
                                   \left( 1 - {3\omega\over 8\tau}   
                                   \right)  \,,\\
k &=& 
-{27\over 5\tau} \left( \sqrt{\beta\tau\over 48} - {\beta\over 24}
                 \right)
                 \left( 1 - {\omega\over 12\tau}   
                 \right)  \,,   
\end{eqnarray}
with $\rho=11+2N_f/N_c^3\,$ and $\omega=1+N_f/N_c^3$. The term 
$\bar{\xi_0}$ is a non-perturbative offset to the $\xi$ distribution.

We have fitted Equation~(\ref{eq:fw}) simultaneously to the \ksip\ 
distribution measured in the present study and in previous studies by 
OPAL~\cite{OPAL-91,OPAL-133,OPAL-161,OPAL-172-189} 
and lower energy experiments~\cite{TASSO-14-44,TOPAZ-58}, covering the 
range $\rs=14-202$~GeV. As the energy scale involved in coherent 
parton emissions, as described by the formalism, is expected to be 
low, we apply the formalism assuming three active flavours. The free 
parameters in the prediction are $\Lambda$ and $\bar{\xi_0}$. The 
overall normalisation $N(\tau)$ is also not predicted in the 
formalism of Equation~(\ref{eq:fw}). We therefore vary the 
normalisation separately at each c.m. energy in the fit.

When comparing the measured \ksip\ distributions to the predicted $\xi$ 
distributions, we neglect the effects of the finite masses of the hadrons 
produced in the final state. For this reason the highest \ksip\ values, 
corresponding to momenta lower than 300~MeV/$c$, are excluded from the fit, 
avoiding the region where such mass effects could play a significant role. 
Moreover, as the formalism of Equation~(\ref{eq:fw}) is not expected to 
be valid at low $\xi$ (high $x$), where logarithms of the form 
$\ln(1-x)$ are important, data below $\ksip=2.0$ are also not considered 
in the fit. 

Figure~\ref{fig:fw} shows the \ksip\ data compared to the fitted 
predictions of Equation~(\ref{eq:fw}). For clarity, not all OPAL data 
included in the fit are shown in the figure, specifically our data at 
161, 172, 183 and 189~GeV. The data in the range included in the fit 
(full lines) are well described by the prediction. At low \ksip\ the 
prediction clearly lies above the data, while at high \ksip\ it appears 
to give a reasonable description of the measurements beyond the fitted range. 
The total \chisq\ of the fit is 121 for 174 degrees of freedom when including 
the full experimental errors (statistical and systematic errors added in 
quadrature) in the fit and assuming them to be uncorrelated. This is the 
only way to treat all data consistently since for the lower energy results 
only the combined statistical and systematic uncertainties are available.
As a consequence, the quoted \chisq\ value presumably underestimates the 
true \chisq\ of the fit.

The obtained values for $\Lambda$ and $\bar{\xi_0}$ are 
$153\pm7\pm55$~MeV and $-0.70\pm0.03\pm0.23$, respectively, where the 
first errors are the fit uncertainties, defined as the square root of the 
diagonal elements of the covariance matrix, and the second errors were 
obtained by varying the fit range. The lower limit on \ksip\ was varied 
between 1.5 and 2.5 and the upper limit between the \ksip\ values 
corresponding to particle momenta of 200 and 400~MeV/$c$. 
The errors on the parameters do not include an estimate of how they may be 
affected by missing terms in the predictions. Therefore these parameters 
are only meaningful within the formalism. A more detailed test of the 
formalism of Equation (\ref{eq:fw}) than the one presented here would need 
to account for more subtle effects such as e.g. the changing 
flavour composition with c.m. energy. For this, additional information
would need to be included in the fit, which goes beyond the scope of our
analysis.

\subsubsection{MLLA prediction}\label{section:mlla}
A prediction for the $\xi$ distribution can also be formulated in the 
so-called modified leading-log-approximation (MLLA, for a review 
see e.g.~\cite{khozeochs}). This approximation constitutes a complete 
resummation of single and double logarithmic terms~\cite{dfk-coh,dfk-dlogs}. 
In the MLLA approach a so-called limiting spectrum prediction for the 
$\xi$ distribution can be formulated in which the QCD scale, $\Lambda$, 
and the scale at which the perturbative process terminates, $Q_0$, are 
taken to have the same value. 
In this study we use the equations given in~\cite{khozeochs}:
\begin{equation}
{1\over \sigma} {\d\sigma^h\over \d\xi} = 
2 K^h {C_F\over N_c} D^{\rm lim}(\xi,\tau)
\label{eq:mlla1},
\end{equation}
where $C_F=(N_c^2-1)/2N_c$ and $K^h$ is a hadronisation constant which 
accounts for the number of hadrons of type $h$ produced per final state 
parton. In some earlier publications the factor of 2 in 
Equation~(\ref{eq:mlla1}), accounting for the two quark jets present 
in $e^+e^-$ annihilation, was incorporated into $K^h$. From the LPHD 
hypothesis $K^h$ is expected to be independent of the underlying process, 
including its scale. For the expression for the limiting spectrum 
distribution, $D^{\rm lim}$, we refer to~\cite{khozeochs}.
In this formalism $\tau=\ln{(\rs/2\Lambda_{\rm eff})}$, where 
$\Lambda_{\rm eff}$ is an effective scale factor representing both the QCD 
scale, $\Lambda$, and the cut-off scale, $Q_0$. Contrary to the Fong and 
Webber prediction, the MLLA limiting spectrum formalism predicts not only 
the shape of the $\xi$ distribution but, apart from the hadronisation 
correction, also its normalisation.

\begin{figure}[t]
\begin{center}
\resizebox{15cm}{!}
{\includegraphics{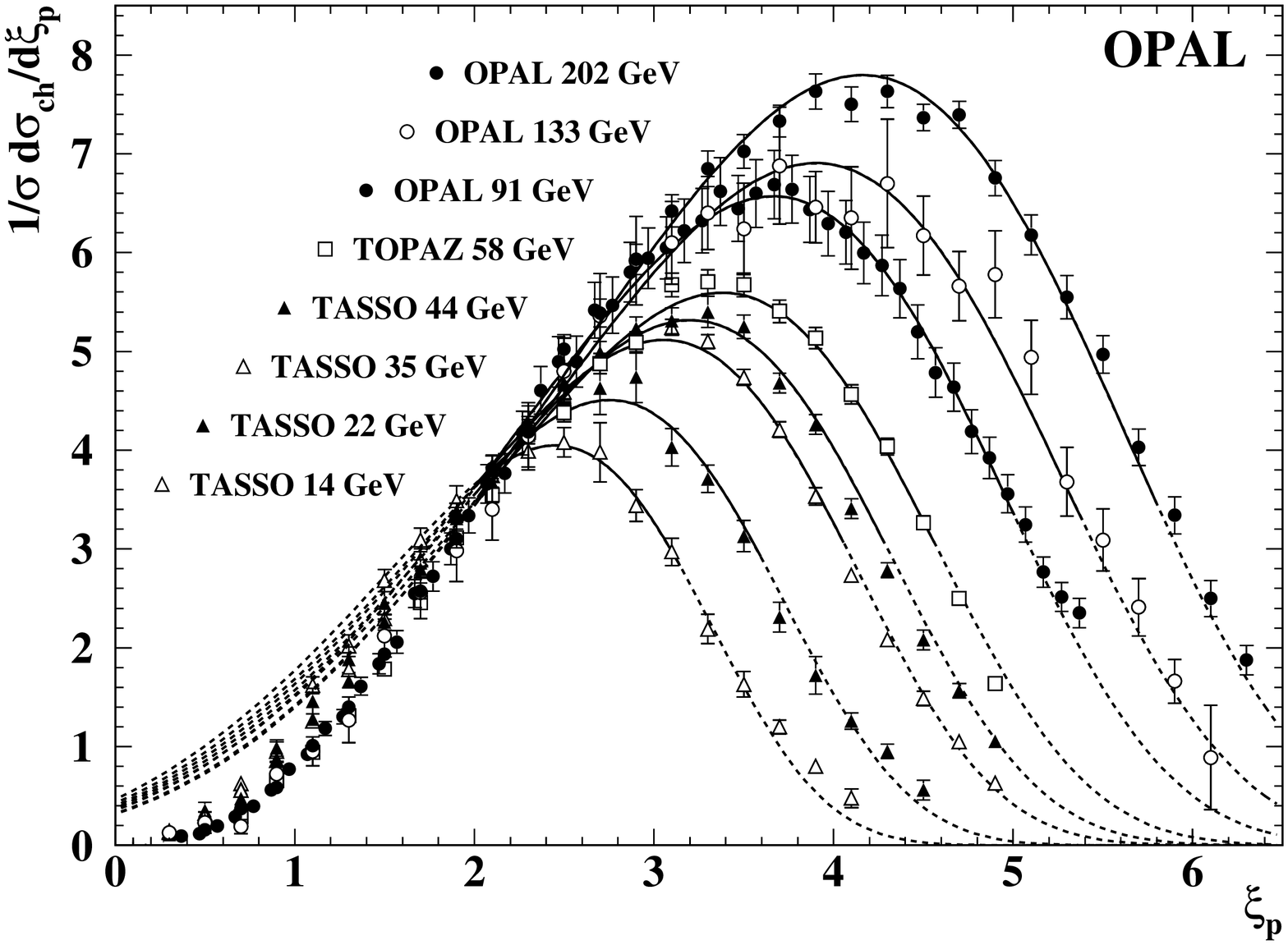}}
\caption[]{ \ksip\ spectra measured in the range $\rs=14-202$~GeV, 
compared to the Fong and Webber~\cite{fongwebber91} predictions 
(Equation (\ref{eq:fw}) in this paper), fitted to the data. The 
full lines indicate the region of the fit. For clarity, not all 
OPAL data included in the fit are shown in the figure. The error 
bars on the data represent the combined statistical and systematic 
uncertainties.}
\label{fig:fw}
\end{center}
\end{figure}

\begin{figure}[t]
\begin{center}
\resizebox{15cm}{!}
{\includegraphics{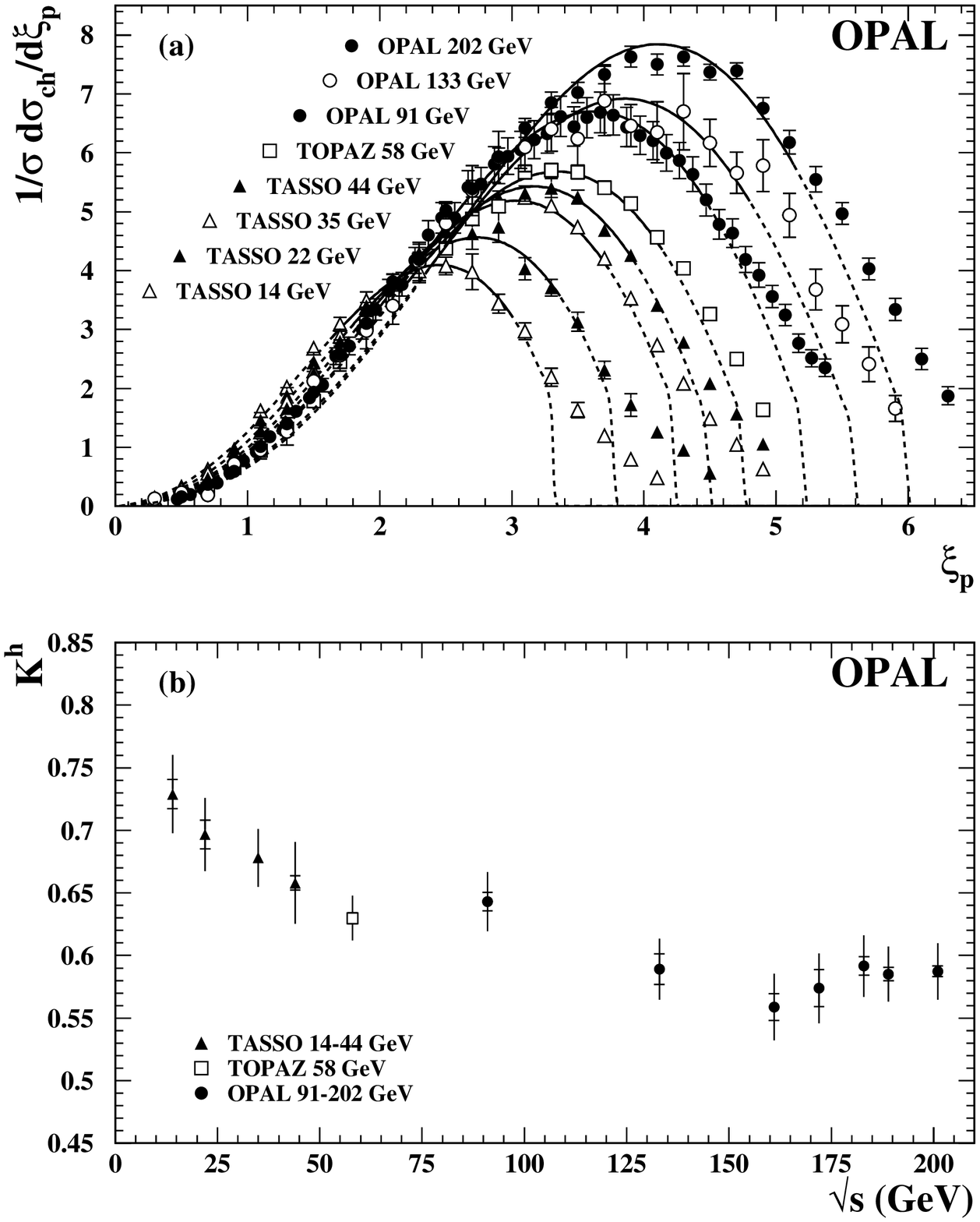}}
\caption[]{ a) \ksip\ spectra measured in the range $\rs=14-202$~GeV, 
compared to MLLA predictions in the limiting spectrum approach 
(\cite{khozeochs}, Equation~(\ref{eq:mlla1}) in this paper), fitted 
to the data. The full lines indicate the region of the fit. For 
clarity, not all OPAL data included in the fit are shown in the figure. 
The error bars on the data represent the combined statistical and 
systematic uncertainties.
  \\
  b) The hadronisation constants $K^h$ at different \rs\ as obtained 
from the fit, described in Section~\ref{section:mlla}. The inner 
error bar represents the fit error and the outer error bar the 
combined fit error and the error obtained from varying the fit range.}
\label{fig:mlla}
\end{center}
\end{figure}

In Figure~\ref{fig:mlla}a, the same \ksip\ data shown in 
Figure~\ref{fig:fw} are compared to the fitted predictions of 
Equation~(\ref{eq:mlla1}), where the free parameters are the effective 
scale, $\Lambda_{\rm eff}$, and a separate hadronisation constant, 
$K^h$, for each c.m. energy. We choose to fit a separate hadronisation 
constant at every energy in order subsequently to examine the obtained 
$K^h$ values to establish how well the formalism and the LPHD hypothesis 
work. Higher order effects, not accounted for in the formalism, are 
expected to give rise to some energy dependence in the $K^h$ factors, 
in particular at the lowest c.m. energies (see e.g.~\cite{dkt91}). As 
in Section~\ref{section:fw}, we apply the formalism assuming three active 
flavours. The fit is restricted to the region around the peaks defined by
$0.75+0.33\log(\rs\,)<\ksip<0.9+0.8\log(\rs\,)$. 

We find the fit describes the data well within the fitted range 
(full line) as illustrated by the total \chisq\ of the fit of 122 for 132 
degrees of freedom. Outside the fit range, both at low and high \ksip, 
deviations of the predictions from the data are observed. 
The MLLA fit is seen to describe the low \ksip\ data better, but the 
high \ksip\ data worse, than the fit of the Fong and Webber prediction
discussed in Section~\ref{section:fw} (see Figure~\ref{fig:fw}).
The value of $\Lambda_{\rm eff}$ obtained from the fit is 
$254\pm 3\pm 31$~MeV, where the first error is the fit uncertainty and 
the second error is obtained from varying the fit range. Both the lower 
and upper limits on \ksip\ were lowered and raised by 0.5 units in \ksip.

In Figure~\ref{fig:mlla}b the obtained $K^h$ factors are shown as a 
function of the c.m. energy. The errors shown are the fit error 
and the combined error from the fit and from varying the fit range, 
as above. Above 130~GeV the $K^h$ factors obtained from our fit are 
consistent with being constant. Below 50~GeV a modest rise in the values of 
the $K^h$ factors is observed. 

\subsubsection{\boldmath The energy dependence of \ksinul }
The MLLA also provides a prediction for the energy evolution of 
the peak position, \ksinul, of the $\xi$ distribution~\cite{dkt92,dkt91}:
\begin{equation}
\xi_0 = \tau \left[   {1\over 2} 
                    + \sqrt{C\over \tau} 
                    - {C\over \tau} 
                    + \mathcal{O}(\tau^{-{3\over 2}}) 
             \right] , 
\label{eq:mllaxi0}
\end{equation}
with $\tau$ defined as in Section~\ref{section:fw} and $C=a^2/16N_c b$.

To determine the peak position \ksinul\ of the measured \ksip\ 
distributions in each of the energy regions of the present study, we 
fitted Equation~(\ref{eq:fw}) to these distributions in the 
range $2.0 < \ksip\ < 6.2$, taking the position of the maximum of the 
fitted function as the result for \ksinul.
The systematic uncertainties are determined by repeating the fit using
the systematic variations described in Section~\ref{section:systematics}. 
In addition, the uncertainty due to the choice 
of the fit range is estimated by repeating the fit using two alternative 
fit ranges: $3.2 < \ksip < 4.8$ and  $1.6 < \ksip < 6.6$, taking the larger 
deviation from the nominal result as the systematic uncertainty. The 
uncertainties are added in quadrature to obtain the total systematic 
uncertainty. As for the momentum distributions, we combine the \ksinul\ 
results from the different energies to obtain: 
\begin{eqnarray}
 \ksinul(201.7~\mathrm{GeV}) & = & 4.158\pm0.007\stat\pm0.036\syst.\nonumber
\end{eqnarray}
  
At c.m. energies below 91~GeV, with the exception of the TOPAZ 
result~\cite{TOPAZ-58}, no \ksinul\ determinations were published. 
Therefore, we determined \ksinul\ values for all energies below 91~GeV  
by fitting Equation~(\ref{eq:fw}) to the published \ksip\ (or \xp) results 
in a narrow region around the peak. This procedure ensures that the 
\ksinul\ results are obtained in a uniform manner over the entire range 
of energies studied. The error on \ksinul\ for the lower energies 
is taken from the variation in the peak position when modifying the fit 
range. For energies above 91~GeV this was found to be the dominant source 
of uncertainty. 

\begin{figure}[t]
\begin{center}
\resizebox{15cm}{!}
{\includegraphics{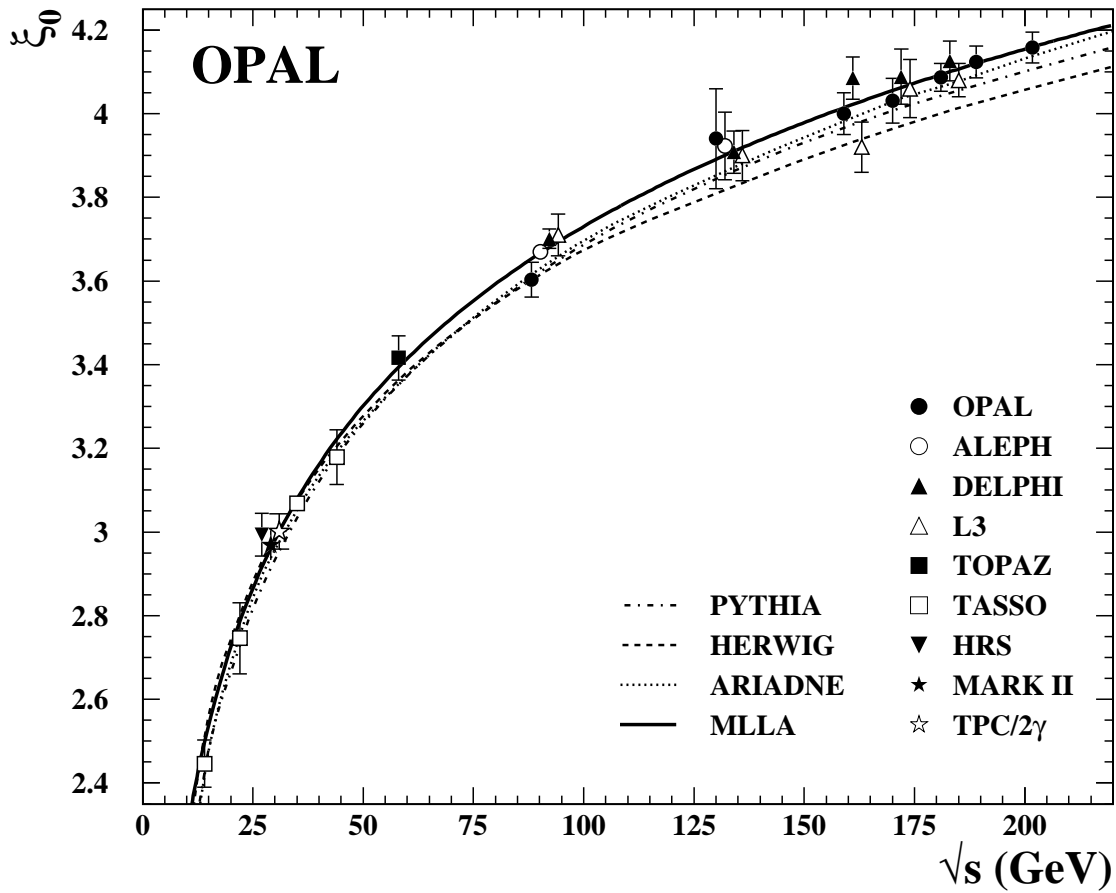}}
\caption[]{ Evolution of the position of the peak in the 
\ksip\ distribution, \ksinul, with c.m. energy, in the range 
$\rs=14-202$~GeV, compared to the fitted MLLA prediction 
(Equation~(\ref{eq:mllaxi0}) in this paper) and to predictions 
of PYTHIA, HERWIG and ARIADNE. The error bars of the data 
represent the combined statistical and systematic uncertainties.}
\label{fig_xi0}
\end{center}
\end{figure}

In Figure~\ref{fig_xi0} our \ksinul\ result is shown together with 
earlier \ksinul\ results from 
OPAL~\cite{OPAL-91,OPAL-133,OPAL-161,OPAL-172-189} 
and other LEP experiments~\cite{L3-133-183,ALEPH-XI0,DELPHI-XI0} and 
the \ksinul\ values we determined for the lower energy 
experiments~\cite{TASSO-14-44,TOPAZ-58,TPC2GAM-29,MARK2-XI0,HRS-29}. 
The results are compared to the predictions of PYTHIA, HERWIG and 
ARIADNE and to the fitted prediction of Equation~(\ref{eq:mllaxi0}). 
For the fit three active flavours were assumed and the scale, 
$\Lambda$, was the only free parameter. The fit yields  
$\Lambda=203\pm2$~MeV, where the error is the fit uncertainty. 
The fitted MLLA prediction is in good agreement with the data. 
The PYTHIA and ARIADNE predictions are slightly below the data, 
while the HERWIG prediction, in particular at the highest c.m. 
energies, is much lower than the data. 

\subsection{\boldmath Comparisons to theory at high $x$}
Another theoretical framework for momentum distributions is 
formulated in terms of the scale evolution of fragmentation 
functions. In this framework, a factorisation between the 
perturbative scattering process on one hand, and the parton 
emission cascade and non-perturbative hadronisation stage on the 
other hand, is formulated as follows 
(see e.g.~\cite{nason,rijken,binnewies}): 
\begin{eqnarray}
{{1\over\sigma} {\d\sigma^h \over \d x}} &=& 
\int_{x}^1 {\d z \over z} 
     \sum_{f=g,u,d,s,c,b} 
           C_f(z,\as,\rs\,)D^h_f({x\over z},\mu).
\label{eq:dsigdx}
\end{eqnarray}
The $C_f(x,\as,\rs)$ are coefficient functions, describing the probability 
to obtain a parton $f$ (gluon, $g$, or quark, $u,d,s,c,b$) with energy 
fraction $x$ in an \epem\ collision, 
and $D^h_f(x,\mu)$ are fragmentation functions, describing the expected 
distribution of final state hadrons $h$ from an initial parton $f$ at 
scale $\mu$. The $x$ dependence of fragmentation functions is not 
known a priori, but their dependence on the energy scale is predicted by 
theory. In QCD this scale dependence, also known as scaling violation, 
is described by parton evolution equations~\cite{dglap}, or 
``DGLAP'' equations. These equations, as well as the coefficient functions 
of Equation~(\ref{eq:dsigdx}), are currently known to next-to-leading 
order (NLO) accuracy in \as~\cite{evol-nlo,nason,rijken,binnewies}. 

Using a computer program~\cite{botje} written for scaling violation 
analyses of proton structure functions in NLO QCD, to which we added NLO 
evolution equations for fragmentation functions and NLO coefficient 
functions for \epem\ annihilation, we performed a fit of 
Equation~(\ref{eq:dsigdx}) to OPAL and lower energy data. These data 
include the inclusive charged particle \xp\ distributions measured here 
and at lower energies~\cite{OPAL-91FLAV,OPAL-133,OPAL-161,OPAL-172-189,
TASSO-14-44,TPC2GAM-29,AMY-55,MARK2-29}, flavour tagged \xp\ 
distributions~\cite{OPAL-91FLAV}, and results on the helicity components 
in charged particle production~\cite{OPAL-91TL}.

The fit involves defining fragmentation functions $D^h_f(x,\mu_0)$ at 
an arbitrarily chosen input scale, $\mu_0$, evolving these over the 
entire relevant range of scales using the parton evolution equations, 
and using them to calculate the theory expectation for 
the measured distributions (Equation~(\ref{eq:dsigdx})). The a priori 
unknown input consists of the input fragmentation functions and the 
strong coupling constant \as. The fragmentation functions are 
parameterised, similarly to reference~\cite{OPAL-91TL}, as
\begin{eqnarray}
xD^h_i(x,\mu_0) &=& N_i(1-x)^{\alpha_i} x^{\beta_i} e^{-c_i(\ln{x})^2}, 
\end{eqnarray}
where the parameters $N_i$, $\alpha_i$, $\beta_i$ and $c_i$ are  
determined independently for $i=g,u,(ds),c,b$. The separation between 
the fragmentation functions of the different quark flavours relies on 
the $(uds)$, $c$ and $b$ flavour tagged data included in the 
fit and on the fact that as a consequence of the different electric and 
weak charges the relative contributions from $d$, $s$ and $b$ quarks 
compared those from $u$ and $c$ quarks, vary with the c.m. energy.  
No data included in the fit provide information to separate the
 contributions from $d$ and $s$ quarks. Therefore only one 
fragmentation function is defined for these two flavours.

In earlier studies~\cite{nason,binnewies,aleph-sv,delphi-sv,kkp-sv}, 
similar fits were performed, including data up to c.m. energies of 91~GeV, 
to obtain measurements of \as. The higher energy LEP~2 data do not add 
significantly to the achievable precision on \as, which is mostly 
constrained by the very precise low energy and LEP~1 results. We 
therefore choose, in our default fit, to fix \as\ at its current 
best estimate of $\asmz=0.1181\pm0.002$~\cite{PDG}, to test how well 
the fitted predictions describe the data over the now extended range 
of scales. The data included in the fit are restricted to \xp\ values 
between 0.07 and 0.80. In addition, the particle momenta were required 
to be greater than 1~GeV/$c$, effectively raising the lower \xp\ limit 
for the lowest energy experiments. This cut is motivated by the fact 
that the present formalism does not account for destructive 
interference effects in soft gluon emissions (see 
Section~\ref{section:low-x}).

\begin{figure}[t]
\begin{center}
\resizebox{15cm}{!}
{\includegraphics{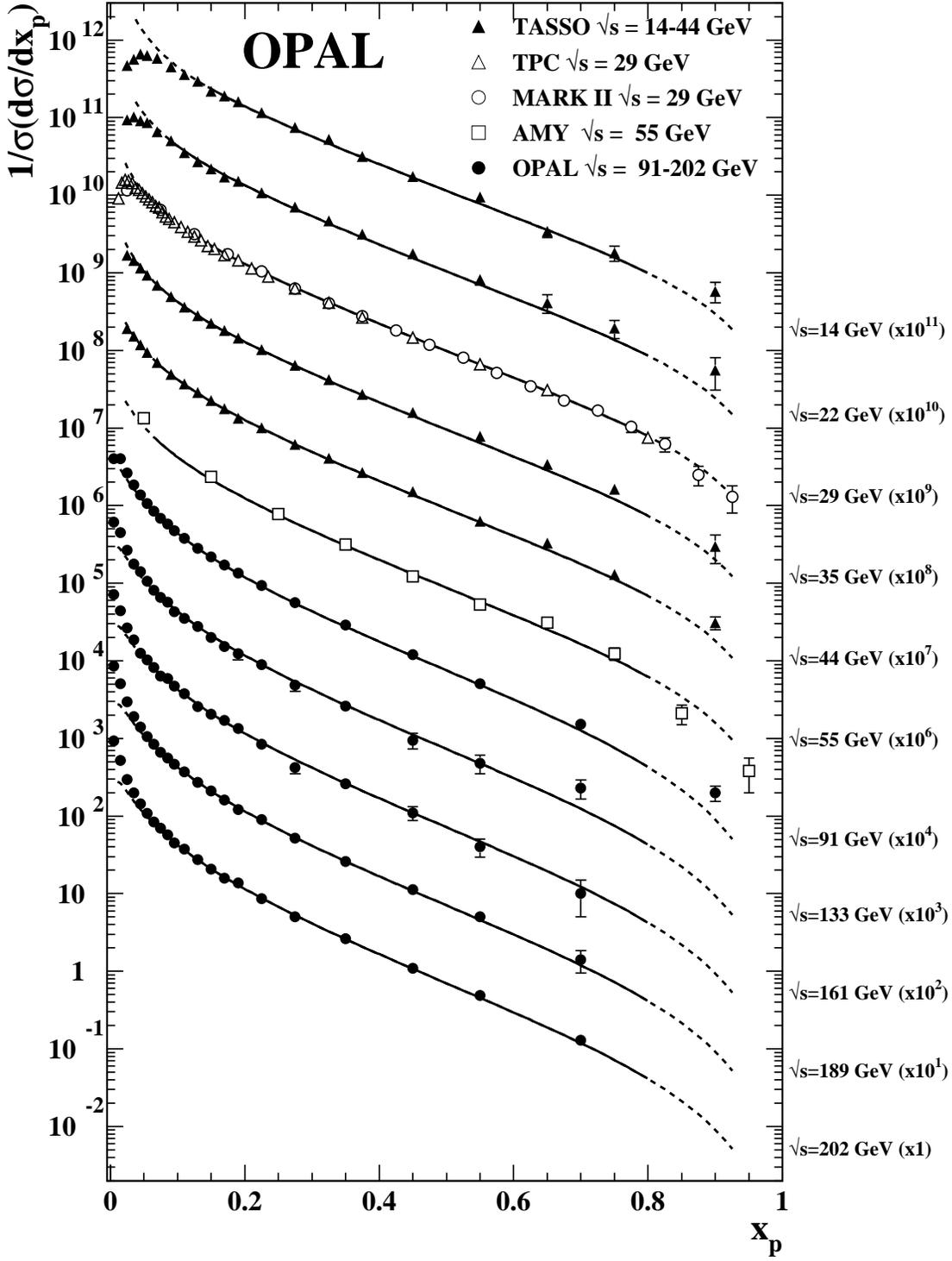}}
\caption[]{ \xp\ spectra measured in the range $\rs=14-202$~GeV, 
compared to fitted NLO predictions (Equation~(\ref{eq:dsigdx}) 
in this paper) using the world average value of $\asmz=0.1181$. 
The full lines indicate the region of the fit. The uncertainties 
of the data represent the combined statistical and systematic 
errors.}
\label{fig:dsigdx}
\end{center}
\end{figure}

In Figure~\ref{fig:dsigdx}, \xp\ distributions measured at c.m. energies 
from 14 to 202~GeV are compared to the fitted predictions. The theory 
predictions are in good agreement with the data in the fitted range (full 
lines). The total \chisq\ of the fit is 198 for 188 degrees of freedom. 
As for the fits described in Section~\ref{section:low-x}, this \chisq\ 
result is based on the full (statistical and systematic) uncertainties on 
the data and neglects any correlations between data points. Outside the 
fitted range, at low \xp\ and low c.m. energies, charged particle 
production is suppressed. This suppression is not described by the 
predictions, which do not include soft gluon interference effects, as 
stated above. 

\begin{figure}[t]
\begin{center}
\resizebox{15cm}{!}
{\includegraphics{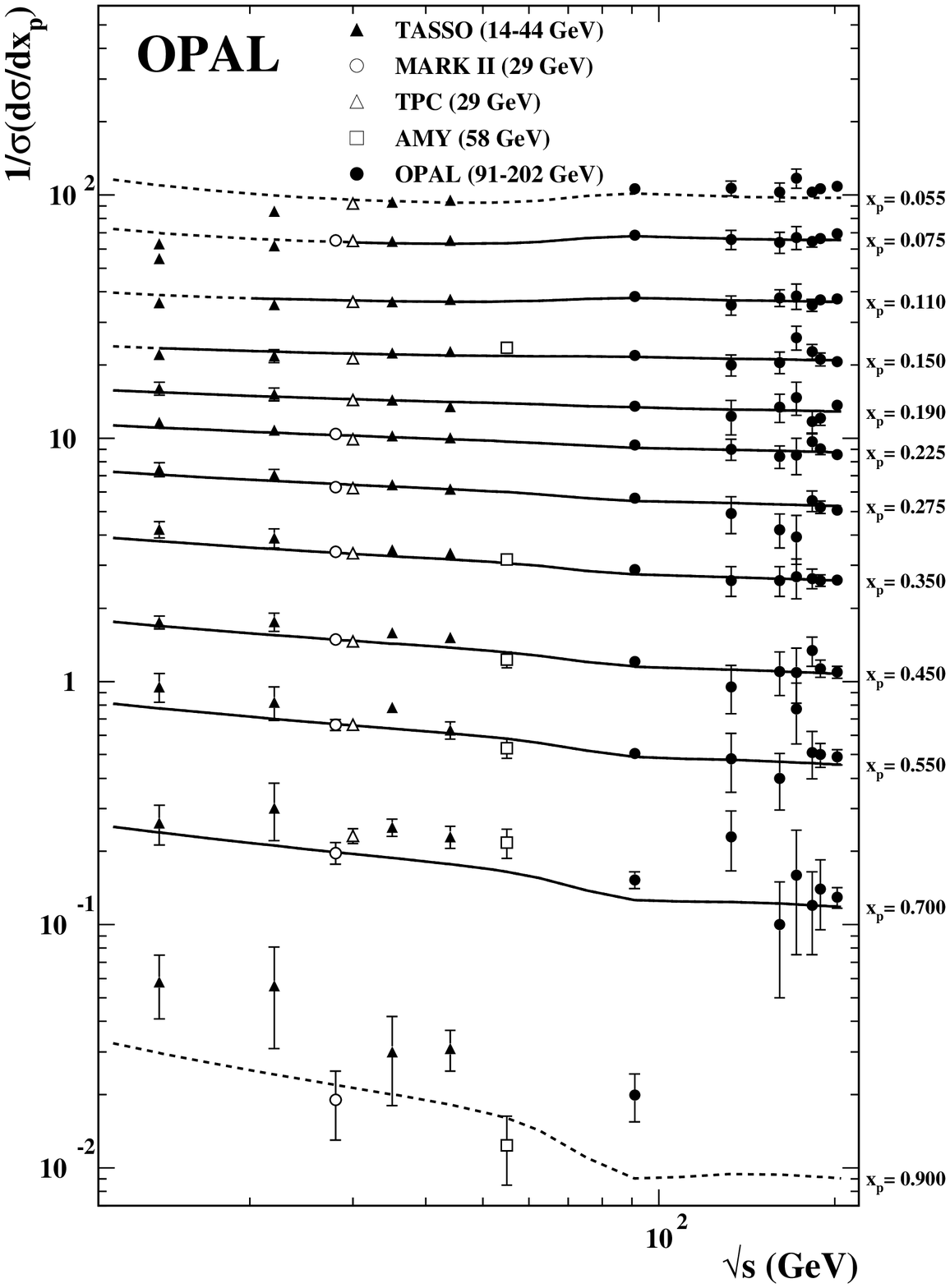}}
\caption[ ]
{ Inclusive charged particle rate, $\sdscd\xp$, as a function of \rs\ in bins of \xp, compared to fitted NLO predictions (Equation~(\ref{eq:dsigdx}) in this paper) using the world average value of $\asmz=0.1181$. The full lines indicate the region of the fit. The uncertainties of the data represent the combined statistical and systematic errors.}
\label{fig:dsigdq}
\end{center}
\end{figure}

Scaling violations in charged particle production are shown in 
Figure~\ref{fig:dsigdq}, where the $\sdscd\xp$ results are shown as 
a function of \rs\ for a number of \xp\ bins. Negative scaling 
violations are visible for the higher \xp\ values. At low \xp\ the 
charged particle production rate demonstrates little dependence on 
the c.m. energy. Changes in the flavour composition with c.m. energy 
also influence the observed level of scaling violations, especially at 
high \xp\ and around $\rs=91$~GeV where a relatively large fraction 
of events is expected to contain $b$ quarks.

The data in Figure~\ref{fig:dsigdq} are also compared to the fitted NLO 
predictions. Overall, a good agreement between data and theory is found 
in the region of the fit (full lines). We thus conclude that inclusive 
charged particle production data can be described by the QCD based 
parton evolution formalism, using the world-average value for \as. 
When we leave \as\ as a free parameter in the fit, we obtain 
$\asmz=0.113\pm0.005\pm0.007$, where the first error is the fit 
uncertainty and the second the uncertainty obtained by independently 
varying the renormalisation and factorisation scales up and down by a 
factor 2. For the dependence of Equation (\ref{eq:dsigdx}) and the 
parton evolution equations on these scales we refer 
to~\cite{dglap,evol-nlo,nason,rijken,binnewies}.

\section{ Summary and conclusion }
\label{section:summary}
We have presented a measurement of charged particle momentum distributions 
in hadronic \epem\ annihilation events produced at LEP~2 at c.m. energies 
between 192 and 209~GeV. The results are combined into a single set of 
distributions,  corresponding to an average c.m. energy of 
$\langle\rs\,\rangle=201.7$ GeV.  

The results, in combination with those obtained at lower c.m. energies, 
were compared to QCD Monte Carlo models and to analytical QCD predictions 
calculated in various approaches. In general, good agreement is 
observed between the data and these theory predictions in the regions 
where the latter are expected to be valid.

\section*{Acknowledgements}
We particularly wish to thank the SL Division for the efficient operation
of the LEP accelerator at all energies
 and for their close cooperation with
our experimental group.  In addition to the support staff at our own
institutions we are pleased to acknowledge the  \\
Department of Energy, USA, \\
National Science Foundation, USA, \\
Particle Physics and Astronomy Research Council, UK, \\
Natural Sciences and Engineering Research Council, Canada, \\
Israel Science Foundation, administered by the Israel
Academy of Science and Humanities, \\
Benoziyo Center for High Energy Physics,\\
Japanese Ministry of Education, Culture, Sports, Science and
Technology (MEXT) and a grant under the MEXT International
Science Research Program,\\
Japanese Society for the Promotion of Science (JSPS),\\
German Israeli Bi-national Science Foundation (GIF), \\
Bundesministerium f\"ur Bildung und Forschung, Germany, \\
National Research Council of Canada, \\
Hungarian Foundation for Scientific Research, OTKA T-029328, 
and T-038240,\\
Fund for Scientific Research, Flanders, F.W.O.-Vlaanderen, Belgium.\\


\pagebreak


\begin{table}[!htb]
{ \begin{center}
    \begin{tabular}{|l|ccc|} 
      \hline 
      \rs\ range (GeV)                   &191 -- 193      &195 -- 197  &199 -- 201    \\
      \hline
      \hline
      $\langle\rs\,\rangle$ (GeV)        &191.6           &195.5       &199.5         \\
      \hline
      Integrated luminosity (\invpb)     &$29.15\pm0.09$&$72.57\pm0.19$&$74.51\pm0.20$\\
      \hline
      {\bf Preselection}                 &{\bf  2796}   &{\bf  6820}   &{\bf  6262}   \\
      Expected                           &$2769\pm8$    &$6448\pm16$   &$6284\pm15$   \\
      4-fermion background (\perc)       &$20.1\pm0.1$  &$21.9\pm0.1$  &$24.8\pm0.1$  \\
      Eff. non-rad. events (\perc)       &$98.4\pm0.1$  &$98.3\pm0.1$  &$98.5\pm0.1$  \\
      \hline
      {\bf ``ISR-fit'' Selection}        &{\bf  765}    &{\bf 1848}    &{\bf 1734}    \\
      Expected                           &$745\pm4$     &$1793\pm8$    &$1771\pm8$    \\
      4-fermion background (\perc)       &$30.2\pm0.3$  &$31.8\pm0.3$  &$33.6\pm0.3$  \\
      Eff. non-rad. events (\perc)       &$89.9\pm0.3$  &$90.0\pm0.2$  &$89.8\pm0.2$  \\
      \hline
      {\bf Final selection}              &{\bf 507}     &{\bf 1111}    &{\bf 1047}    \\
      Expected                           &$461\pm4$     &$1074\pm6$    &$1039\pm6$    \\
      4-fermion background (\perc)       &$4.6\pm0.1$   &$4.6\pm0.1$   &$5.3\pm0.1$   \\
      Eff. non-rad. events (\perc)       &$77.0\pm0.4$  &$76.3\pm0.3$  &$76.1\pm0.3$  \\
\hline 
\multicolumn{4}{}{                         }\\
\multicolumn{4}{}{                         }\\
\multicolumn{4}{}{                         }\\
\hline 
      \rs\ range (GeV)                   &201 -- 202.5  &202.5 -- 205.5&205.5 -- 209.5 \\ 
      \hline
      \hline
      $\langle\rs\,\rangle$ (GeV)        &201.6         &204.9         &206.6     \\ 
      \hline
      Integrated luminosity (\invpb)     &$37.97\pm0.11$&$81.96\pm0.19$&$137.14\pm0.29$\\ 
      \hline
      {\bf Preselection}                 &{\bf  2971}   &{\bf  6031}   &{\bf  9847}\\ 
      Expected                           &$3021\pm8$    &$6240\pm16$   &$10114\pm26$ \\ 
      4-fermion background (\perc)       &$24.4\pm0.1$  &$25.8\pm0.2$  &$26.5\pm0.2$\\
      Eff. non-rad. events (\perc)       &$98.3\pm0.1$  &$98.5\pm0.1$  &$98.4\pm0.1$\\ 
      \hline
      {\bf ``ISR-fit'' Selection}        &{\bf  838}    &{\bf 1781}    &{\bf 2840}\\
      Expected                           &$881\pm4$     &$1857\pm8$    &$3052\pm14$  \\ 
      4-fermion background (\perc)       &$33.7\pm0.3$  &$34.5\pm0.3$  &$34.8\pm0.3$\\
      Eff. non-rad. events (\perc)       &$89.4\pm0.2$  &$89.8\pm0.2$  &$89.7\pm0.2$\\ 
      \hline
      {\bf Final selection}              &{\bf  494}    &{\bf 1091}    &{\bf 1647}\\ 
      Expected                           &$514\pm3$     &$1075\pm6$    &$1758\pm10$  \\ 
      4-fermion background (\perc)       &$5.2\pm0.1$   &$5.6\pm0.2$   &$5.7\pm0.2$ \\
      Eff. non-rad. events (\perc)       &$75.6\pm0.3$  &$75.8\pm0.3$  &$75.9\pm0.3$\\ 
\hline 
    \end{tabular}
  \end{center}
  }
  \caption[]{ Data samples corresponding to six ranges in c.m. energy. 
The luminosities and the luminosity weighted mean c.m. energies are 
given in the first two rows. The rows labelled as ``Preselection'', 
``ISR-fit selection'' and ``Final selection'' correspond to the number 
of events passing these criteria. For each of these selection stages 
the expected number of events, the remaining fraction of 4-fermion 
background and the signal efficiency are also given. The errors on the 
integrated luminosity include statistical and systematic uncertainties 
added in quadrature. All other errors are the statistical errors only.}
  \label{tab:select}
\end{table}


\begin{table}[t]
 \begin{center}
 \begin{tabular}{|c| |r@{ $\pm$ }l@{ $\pm$ }l|}
 \hline
 \ptin\ (GeV/$c$) &\multicolumn{3}{c|}{\sdscd\ptin\ $({\rm GeV}/c)^{-1}$ }\\
 \hline
 \hline
0.00-0.10 
 &   55.8&   0.4&   2.0
 \\
0.10-0.20 
 &   44.8&   0.3&   1.4
 \\
0.20-0.30 
 &  33.58&  0.29&  0.80
 \\
0.30-0.40 
 &  25.35&  0.24&  0.42
 \\
0.40-0.50 
 &  18.73&  0.21&  0.18
 \\
0.50-0.60 
 &  14.83&  0.19&  0.10
 \\
0.60-0.70 
 &  11.49&  0.17&  0.15
 \\
0.70-0.80 
 &   9.48&  0.15&  0.16
 \\
0.80-0.90 
 &   7.71&  0.14&  0.16
 \\
0.90-1.00 
 &   6.46&  0.13&  0.07
 \\
1.00-1.20 
 &  4.926& 0.083& 0.048
 \\
1.20-1.40 
 &  3.689& 0.073& 0.038
 \\
1.40-1.60 
 &  2.933& 0.064& 0.057
 \\
1.60-2.00 
 &  2.052& 0.043& 0.054
 \\
2.00-2.50 
 &  1.240& 0.030& 0.045
 \\
2.50-3.00 
 &  0.754& 0.023& 0.019
 \\
3.00-3.50 
 &  0.528& 0.020& 0.023
 \\
3.50-4.00 
 &  0.360& 0.016& 0.020
 \\
4.00-5.00 
 &  0.217& 0.010& 0.011
 \\
5.00-6.00 
 &  0.123& 0.008& 0.007
 \\
6.00-7.00 
 &  0.074& 0.006& 0.005
 \\
7.00-8.00 
 &  0.054& 0.005& 0.006
 \\
 \hline
 \end{tabular}
 \end{center}
\caption[]{ Measured values of the distribution \sdscd\ptin\ at 
$\langle\rs\,\rangle = 201.7$~GeV. The first error is statistical, 
the second systematic.}
\label{tab:pi}
\end{table}

\begin{table}[t]
 \begin{center}
 \begin{tabular}{|c| |r@{ $\pm$ }l@{ $\pm$ }l|}
 \hline
 \ptout\ (GeV/$c$)&\multicolumn{3}{c|}{\sdscd\ptout\ $({\rm GeV}/c)^{-1}$}\\
 \hline
 \hline
0.00-0.10 
 &   81.7&   0.5&   2.0
 \\
0.10-0.20 
 &   62.2&   0.4&   1.2
 \\
0.20-0.30 
 &  43.39&  0.33&  0.46
 \\
0.30-0.40 
 &  28.93&  0.27&  0.31
 \\
0.40-0.50 
 &  19.14&  0.22&  0.25
 \\
0.50-0.60 
 &  12.60&  0.18&  0.29
 \\
0.60-0.70 
 &   8.49&  0.15&  0.21
 \\
0.70-0.80 
 &   6.08&  0.13&  0.17
 \\
0.80-0.90 
 &   4.14&  0.11&  0.07
 \\
0.90-1.00 
 &  3.143& 0.098& 0.054
 \\
1.00-1.20 
 &  1.883& 0.059& 0.072
 \\
1.20-1.40 
 &  1.266& 0.049& 0.064
 \\
1.40-1.60 
 &  0.798& 0.040& 0.042
 \\
1.60-2.00 
 &  0.445& 0.024& 0.017
 \\
2.00-2.40 
 &  0.237& 0.020& 0.025
 \\
2.40-2.80 
 &  0.102& 0.014& 0.034
 \\
2.80-3.20 
 &  0.047& 0.011& 0.018
 \\
3.20-3.60 
 &  0.035& 0.011& 0.026
 \\
 \hline
 \end{tabular}
 \end{center}
\caption[]{ Measured values of the distribution \sdscd\ptout\ at 
$\langle\rs\,\rangle = 201.7$~GeV. The first error is statistical, 
the second systematic.}
\label{tab:po}
\end{table}

\begin{table}[t]
 \begin{center}
 \begin{tabular}{|c| |r@{ $\pm$ }l@{ $\pm$ }l|}
 \hline
 \yp            &\multicolumn{3}{c|}{\sdscd\yp                         }\\
 \hline
 \hline
0.00-0.33 
 &   7.44&  0.12&  0.34
 \\
0.33-0.67 
 &   8.26&  0.13&  0.34
 \\
0.67-1.00 
 &   8.22&  0.12&  0.25
 \\
1.00-1.33 
 &   8.26&  0.11&  0.15
 \\
1.33-1.67 
 &   8.08&  0.10&  0.12
 \\
1.67-2.00 
 &  7.65& 0.09& 0.09
 \\
2.00-2.33 
 &  7.18& 0.08& 0.10
 \\
2.33-2.67 
 &   6.56&  0.07&  0.12
 \\
2.67-3.00 
 &   5.91&  0.06&  0.15
 \\
3.00-3.33 
 &   5.12&  0.06&  0.17
 \\
3.33-3.67 
 &   4.14&  0.05&  0.14
 \\
3.67-4.00 
 &  3.034& 0.047& 0.088
 \\
4.00-4.33 
 &  1.936& 0.039& 0.050
 \\
4.33-4.67 
 &  1.118& 0.029& 0.022
 \\
4.67-5.00 
 &  0.588& 0.021& 0.014
 \\
5.00-5.33 
 &  0.258& 0.013& 0.014
 \\
5.33-5.67 
 &  0.108& 0.009& 0.008
 \\
5.67-6.00 
 &  0.049& 0.006& 0.003
 \\
6.00-6.33 
 &  0.011& 0.003& 0.001
 \\
 \hline
 \end{tabular}
 \end{center}
\caption[]{ Measured values of the distribution \sdscd\yp\ at 
$\langle\rs\,\rangle = 201.7$~GeV. The first error is statistical, 
the second systematic.}
\label{tab:yp}
\end{table}

\begin{table}[t]
 \begin{center}
 \begin{tabular}{|c| |r@{ $\pm$ }l@{ $\pm$ }l|}
 \hline
 \xp            &\multicolumn{3}{c|}{\sdscd\xp                         }\\
 \hline
 \hline
0.00-0.01 
 &   927.&    7.&   23.
 \\
0.01-0.02 
 &   522.&    4.&   10.
 \\
0.02-0.03 
 &  296.4&   2.8&   6.0
 \\
0.03-0.04 
 &  199.8&   2.2&   3.9
 \\
0.04-0.05 
 &  144.2&   1.9&   3.3
 \\
0.05-0.06 
 &  108.4&   1.6&   2.3
 \\
0.06-0.07 
 &   84.0&   1.3&   2.2
 \\
0.07-0.08 
 &   69.4&   1.2&   1.5
 \\
0.08-0.09 
 &   57.6&   1.1&   1.4
 \\
0.09-0.10 
 &   45.2&   1.0&   1.0
 \\
0.10-0.12 
 &   37.4&   0.6&   1.1
 \\
0.12-0.14 
 &  27.59&  0.55&  0.72
 \\
0.14-0.16 
 &  20.60&  0.47&  0.69
 \\
0.16-0.18 
 &  15.76&  0.42&  0.35
 \\
0.18-0.20 
 &  13.69&  0.39&  0.34
 \\
0.20-0.25 
 &   8.57&  0.19&  0.17
 \\
0.25-0.30 
 &   5.06&  0.15&  0.17
 \\
0.30-0.40 
 &   2.61&  0.07&  0.11
 \\
0.40-0.50 
 &  1.092& 0.045& 0.041
 \\
0.50-0.60 
 &  0.489& 0.029& 0.019
 \\
0.60-0.80 
 &  0.130& 0.009& 0.008
 \\
 \hline
 \end{tabular}
 \end{center}
\caption[]{ Measured values of the distribution \sdscd\xp\ at 
$\langle\rs\,\rangle = 201.7$~GeV. The first error is statistical, 
the second systematic.}
\label{tab:xp}
\end{table}

\begin{table}[t]
 \begin{center}
 \begin{tabular}{|c| |r@{ $\pm$ }l@{ $\pm$ }l|}
 \hline
 \ksip          &\multicolumn{3}{c|}{\sdscd\ksip                       }\\
 \hline
 \hline
0.00-0.20 
 &  0.015& 0.003& 0.006
 \\
0.20-0.40 
 &  0.063& 0.006& 0.016
 \\
0.40-0.60 
 &  0.157& 0.011& 0.014
 \\
0.60-0.80 
 &  0.370& 0.018& 0.018
 \\
0.80-1.00 
 &  0.588& 0.024& 0.029
 \\
1.00-1.20 
 &  1.012& 0.033& 0.043
 \\
1.20-1.40 
 &  1.398& 0.039& 0.047
 \\
1.40-1.60 
 &  1.936& 0.045& 0.044
 \\
1.60-1.80 
 &  2.575& 0.053& 0.060
 \\
1.80-2.00 
 &  3.106& 0.058& 0.075
 \\
2.00-2.20 
 &   3.82&  0.06&  0.11
 \\
2.20-2.40 
 &   4.19&  0.07&  0.11
 \\
2.40-2.60 
 &   5.02&  0.07&  0.12
 \\
2.60-2.80 
 &   5.39&  0.08&  0.12
 \\
2.80-3.00 
 &   5.93&  0.08&  0.13
 \\
3.00-3.20 
 &   6.42&  0.09&  0.14
 \\
3.20-3.40 
 &   6.85&  0.09&  0.15
 \\
3.40-3.60 
 &   7.02&  0.09&  0.15
 \\
3.60-3.80 
 &   7.33&  0.09&  0.13
 \\
3.80-4.00 
 &   7.63&  0.10&  0.15
 \\
4.00-4.20 
 &   7.50&  0.10&  0.14
 \\
4.20-4.40 
 &   7.63&  0.10&  0.13
 \\
4.40-4.60 
 &  7.37& 0.10& 0.09
 \\
4.60-4.80 
 &  7.39& 0.10& 0.09
 \\
4.80-5.00 
 &   6.75&  0.09&  0.15
 \\
5.00-5.20 
 &   6.18&  0.09&  0.18
 \\
5.20-5.40 
 &   5.55&  0.08&  0.20
 \\
5.40-5.60 
 &   4.97&  0.08&  0.17
 \\
5.60-5.80 
 &   4.03&  0.07&  0.17
 \\
5.80-6.00 
 &   3.34&  0.06&  0.18
 \\
6.00-6.20 
 &   2.50&  0.06&  0.17
 \\
 \hline
 \end{tabular}
 \end{center}
\caption[]{ Measured values of the distribution \sdscd\ksip\ at 
$\langle\rs\,\rangle = 201.7$~GeV. The first error is statistical, 
the second systematic.}
\label{tab:xi}
\end{table}

\pagebreak


\begin{thebibliography}{99}

\bibitem{OPAL-133} 
 OPAL Collaboration, G.~Alexander et~al., Z. Phys. {\bf C72} (1996) 191
\bibitem{OPAL-161} 
 OPAL Collaboration, K.~Ackerstaff et~al., Z. Phys. {\bf C75} (1997) 193
\bibitem{OPAL-172-189} 
 OPAL Collaboration, G.~Abbiendi et~al., Eur. Phys. J. {\bf C16} (2000) 185
\bibitem{ALEPH-133} 
 ALEPH Collaboration, D.~Busculic et~al., Z. Phys. {\bf C73} (1997) 409
\bibitem{L3-133-183} 
 L3 Collaboration, M.~Acciarri et~al., Phys. Lett. {\bf B444} (1998) 569
\bibitem{DELPHI-133} 
 DELPHI Collaboration, P.~Abreu et~al., Z. Phys. {\bf C73} (1997) 229

\bibitem{opaldetector} 
 OPAL Collaboration, K.~Ahmet et~al., Nucl. Instrum. Methods {\bf A305} (1991) 275

\bibitem{gopal} 
 J.~Allison et~al., Nucl. Instrum. Methods {\bf A317} (1992) 47
\bibitem{jetset74} 
 T.~Sj{\"o}strand, Comput. Phys. Commun. {\bf 82} (1994) 74
\bibitem{pythia61} 
 T.~Sj{\"o}strand et~al., Comput. Phys. Commun. {\bf 135} (2001) 238 
\bibitem{kk2f} 
 S.~Jadach, B.F.L.~Ward and Z.~W\c{a}s, Comput. Phys. Commun. {\bf 130} (2000) 260  
\bibitem{opaltune} 
 OPAL Collaboration, G.~Alexander et al., Z. Phys. {\bf C69} (1996) 543
\bibitem{herwig} 
 G.~Marchesini et~al., Comput. Phys. Commun. {\bf 67} (1992) 465  
\bibitem{grc4f} 
 J.~Fujimoto et~al., Comput. Phys. Commun. {\bf 100} (1997) 128
\bibitem{vermaseren} 
 J.A.M.~Vermaseren, J.~Smith and G.~Grammer Jr., Phys. Rev. {\bf D19} (1979) 137;\\
 J.A.M.~Vermaseren, Nucl. Phys. {\bf B229} (1983) 347
\bibitem{PHOJET} 
 R.~Engel, Z. Phys. {\bf C66} (1995) 203;\\                       
 R.~Engel and J. Ranft, Phys. Rev. {\bf D54} (1996) 4244
\bibitem{Tauola} 
 S.~Jadach et al., Comput. Phys. Commun. {\bf 76} (1993) 361
\bibitem{ariadne} 
 L.~L{\"o}nnblad, Comput. Phys. Commun. {\bf 71} (1992) 15        v4
\bibitem{artune} 
 ALEPH Collaboration, R.~Barate et al., Phys. Rept. {\bf 294} (1998) 1;\\   
 OPAL Collaboration, G.~Abbiendi et al., Eur. Phys. J. {\bf C11} (1999) 217               

\bibitem{TKMH} 
 OPAL Collaboration, G.~Alexander et al.,  Z. Phys. {\bf C52} (1991) 175
\bibitem{OPALPR183} 
 OPAL Collaboration, K.~Ackerstaff et al., Phys. Lett. {\bf B391} (1997) 221
\bibitem{durham} 
 S.~Catani et al., Phys. Lett. {\bf B269} (1991) 432
\bibitem{MT} 
 OPAL Collaboration, K.~Ackerstaff et al., Eur. Phys. J. {\bf C2} (1998) 213
\bibitem{event2} 
 S.~Catani and M.H.~Seymour, Phys. Lett. {\bf B378} (1996) 287
\bibitem{ERT} 
 R.K.~Ellis, D.A.~Ross and A.E.~Terrano, Nucl. Phys. {\bf B178} (1981) 421  

\bibitem{OPALPR321} 
 OPAL Collaboration, G.~Abbiendi et al., Phys. Lett. {\bf B493} (2000) 249


\bibitem{lphd} 
 Ya.I.~Azimov et al., Z. Phys. {\bf C27} (1985) 65;\\
 Ya.I.~Azimov et al., Z. Phys. {\bf C31} (1986) 213     
\bibitem{khozeochs} 
 V.A.~Khoze and W.~Ochs, Int. J. of Mod. Phys. {\bf A12} (1997) 2949
\bibitem{dfk-coh} 
 Yu.L.~Dokshitzer, V.S.~Fadin and V.A.~Khoze, Phys. Lett. {\bf B115} (1982) 242
\bibitem{fongwebber91} 
 C.P.~Fong and B.R.~Webber, Nucl. Phys. {\bf B355} (1991) 54
\bibitem{OPAL-91} 
 OPAL Collaboration, M.Z.~Akrawy et al., Phys. Lett. {\bf B247} (1990) 617
\bibitem{TASSO-14-44} 
 TASSO Collaboration, W.~Braunschweig et al., Z. Phys. {\bf C47} (1990) 187
\bibitem{TOPAZ-58} 
 TOPAZ Collaboration, R.~Itoh et al., Phys. Lett. {\bf B345} (1995) 335

\bibitem{dfk-dlogs} 
 Yu.L.~Dokshitzer, V.S.~Fadin and V.A.~Khoze, Z. Phys. {\bf C15} (1982) 325
\bibitem{dkt91} 
 Yu.L.~Dokshitzer, V.A.~Khoze and S.I.~Troyan, J. Phys. {\bf G17} (1991) 1481

\bibitem{dkt92} 
 Yu.L.~Dokshitzer, V.A.~Khoze and S.I.~Troyan, Int. J. Mod. Phys. {\bf A7} (1992) 1875
\bibitem{ALEPH-XI0} 
 ALEPH Collaboration, R.~Barate et al., Phys. Rept. {\bf 294} (1998) 1
\bibitem{DELPHI-XI0} 
 DELPHI Collaboration, P.~Abreu et al., Phys. Lett. {\bf B459} (1999) 397
\bibitem{TPC2GAM-29} 
 TPC/2$\gamma$ Collaboration, H.~Aihara et al., Phys. Rev. Lett. {\bf 61} (1988) 1263;\\  
 G.D.~Cowan,  Ph.D. Thesis (LBL, Berkeley) LBL-24715 (1988)
\bibitem{MARK2-XI0} 
 MARK II Collaboration, J.F.~Patrick et al., Phys. Rev. Lett. {\bf 49} (1982) 1232
\bibitem{HRS-29} 
 HRS Collaboration, D.~Bender et al., Phys. Rev. {\bf D31} (1985) 1

\bibitem{nason} 
 P.~Nason and B.R.~Webber, Nucl. Phys. {\bf B421} (1994) 473
\bibitem{rijken} 
 P.J.~Rijken and W.L.~van~Neerven, Phys. Lett. {\bf B386} (1996) 422;\\
 P.J.~Rijken and W.L.~van~Neerven, Nucl. Phys. {\bf B487} (1997) 233
\bibitem{binnewies} 
 J.~Binnewies, Ph.D. Thesis (Hamburg U.) DESY 97-128 (1997)
\bibitem{dglap} 
 V.N.~Gribov and L.N.~Lipatov, Sov. J. Nucl. Phys. {\bf 15} (1972) 438;\\
 G.~Altarelli and G.~Parisi, Nucl. Phys. {\bf B126} (1977) 298;\\
 Yu.L.~Dokshitzer, Sov. Phys. JETP {\bf 46} (1977)
\bibitem{evol-nlo} 
 W.~Furmanski and R.~Petronzio, Phys. Lett. {\bf B97} (1980) 437;\\
 G.~Curci, W.~Furmanski and R.~Petronzio, Nucl. Phys. {\bf B175} (1980) 27;\\
 W.~Furmanski and R.~Petronzio, Z. Phys. {\bf C11} (1982) 293
\bibitem{botje} 
 M.~Botje, ``QCDNUM version 16.12'', ZEUS-97-066 (unpublished), see\\
 http,//www.nikhef.nl/user/h24/qcdnum
\bibitem{AMY-55} 
 AMY Collaboration, Y.K.~Li et al., Phys. Rev. {\bf D41} (1990) 2675
\bibitem{MARK2-29} 
 MARK II Collaboration, A.~Petersen et al., Phys. Rev. {\bf D37} (1988) 1
\bibitem{OPAL-91FLAV} 
 OPAL Collaboration, K.~Ackerstaff et al., Eur. Phys. J. {\bf C7} (1999) 369
\bibitem{OPAL-91TL} 
 OPAL Collaboration, R.~Akers et al., Z. Phys. {\bf C68} (1995) 203
\bibitem{aleph-sv} 
 ALEPH Collaboration, D.~Buskulic et al., Phys. Lett. {\bf B357} (1995) 487
\bibitem{delphi-sv} 
 DELPHI Collaboration, P.~Abreu, Phys. Lett. {\bf B398} (1997) 194
\bibitem{kkp-sv} 
 B.A.~Kniehl, G.~Kramer and B.~P\"{o}tter, Nucl. Phys. {\bf B582} (2000) 514
\bibitem{PDG} 
 D.E.~Groom et al., Eur. Phys. J. {\bf C15} (2000) 1                

\end{thebibliography}
\end{document}